\UseRawInputEncoding
\documentclass[aps,prb,twocolumn,superscriptaddress]{revtex4-1}
\usepackage{graphicx,bm,amssymb,amsmath,xcolor,pifont,soul}
\usepackage[linktocpage=true,colorlinks=true,pdfborder={0 0 0},linkcolor=blue,citecolor=red,filecolor=yellow,urlcolor=blue,bookmarks,pdfauthor={},]{hyperref}
\usepackage{ulem}
\usepackage{multirow}

\begin{document}

\title{Functionalized Cr$_2$C  MXenes: Novel Magnetic Semiconductors} 
\author{Yogendra Limbu}
\affiliation{Department of Physics and Astronomy, University of Iowa, Iowa City, Iowa 52242, USA}
\author{Hari Paudyal}
\affiliation{Department of Physics and Astronomy, University of Iowa, Iowa City, Iowa 52242, USA}
\author{Eudes Gomes da Silva}
\affiliation{Department of Physics and Astronomy, University of Iowa, Iowa City, Iowa 52242, USA}
\author{Denis R. Candido}
\affiliation{Department of Physics and Astronomy, University of Iowa, Iowa City, Iowa 52242, USA}
\author{Michael E. Flatt\'e}
\affiliation{Department of Physics and Astronomy, University of Iowa, Iowa City, Iowa 52242, USA}
\affiliation{Department of Applied Physics, Eindhoven University of Technology, Eindhoven, The Netherlands}
\author{Durga Paudyal}
\email{durga.quantum@gmail.com}
\affiliation{Department of Physics and Astronomy, University of Iowa, Iowa City, Iowa 52242, USA}

\begin{abstract}
\noindent We report an \textit{ab initio} investigation of functionalized and 3$d$-electrons doped Cr$_2$C MXenes. 
Upon functionalization, the Cr$_2$C becomes chemically, dynamically, and mechanically stable, and it exhibits magnetic semiconducting behavior. Cr$_2$CF$_2$ stands out as a wide band gap semiconductor, possessing super exchange interaction mediated by F atoms within the layer, however, the applied strain transforms it from an indirect to a direct band gap semiconductor. Strong spin-phonon coupling found in Cr$_2$CH$_2$ is supported by the distorted Cr spin density due to hydrogen environment. Two magnon branches, associated with two sub-lattice spins, are found in the ferromagnetic Cr$_2$CO$_2$ and antiferromagnetic Cr$_2$CF$_2$. Depending on the types of 3$d$-electron dopants and functionalization, Cr$_2$C MXenes (except for Cr$_2$CO$_2$) change from the indirect band gap magnetic semiconductor to  different states of electronic and magnetic matter including exotic direct band gap magnetic bipolar semiconductor. In addition, we reveal a band inversion between the two highest valence bands in the Fe-doped  Cr$_2$CCl$_2$.

\end{abstract}
\maketitle

\noindent The discovery of graphene   opened a new route of uncovering two dimensional (2D)  MXenes \cite{anasori20172d} exhibiting  diverse exotic electronic and magnetic properties. The electronic properties include semi-metallic, semiconducting, superconducting, and topologically insulating~\cite{limbu2022electronic, huang2020large, bekaert2020first, khazaei2016topological}, whereas magnetic properties include unusual ferromagnetic (FM), ferrimagnetic (FIM), and antiferromagnetic (AFM) ~\cite{bae2020materials, xie2013hybrid} behaviors. MXenes exhibit structural and magnetic complexities \cite{si2015half}, leading to a wide range  of applications, e.g., energy storage \cite{sun2018two}, super-capacitors \cite{chaudhari2017mxene}, biomedical\cite{huang2018two},  transparent conductors \cite{mariano2016solution}, field effect transistors \cite{lai2015surface}, hybrid nanocomposites \cite{xue2017preparation}, sensing devices \cite{yu2015monolayer}, nanoscale superconductivity \cite{zhang2017superconductivity}, and spintronic devices \cite{kumar2017tunable}. 

In general, MXene is synthesized by eliminating A atoms from the three dimensional MAX phase~\cite{magnuson2006electronic} and is inhabited with  functional groups, e.g., -O, -OH, -H, -F, and -Cl~\cite{tan2020first, mashtalir2013intercalation}.  The functionalization   acts as a chemical dopant controlling the electronic and magnetic properties~\cite{frey2019surface}, whereas site substitution and mechanical strain act as a physical control influencing the nano-magnetism~\cite{wu2015atomically,hu2019vacancy, zhao2014manipulation}. Due to the involvement of $3d$ electrons and spin orbit coupling (SOC),  MXenes exhibit unique magnetic and topological phenomena, e.g., high Curie temperature ferromagnet~\cite{kumar2017tunable} and time reversal symmetry breaking ~\cite{weng2015large} driven nontrivial band topology \cite{xu2015large,hantanasirisakul2018electronic, fashandi2015dirac}.

 Bipolar magnetic semiconductors, also exhibited by MXenes,  are characterized by the valance and conduction bands of opposite spins  \cite{he2019cr} and controlled by external electric field \cite{li2013bipolar}. These materials are unique to generate and manipulate carriers with total spin polarization and also design magnon-based spintronic devices~\cite{qin2015long, chumak2014magnon}. Magnons are used to transfer the information without any dissipation ~\cite{balashov2014magnon}, and their dispersion is calculated from exchange interactions~\cite{szilva2013interatomic, pajda2001ab}. Additionally, the spin phonon coupling in 2D materials is important due to its diverse application in spintronics and quantum information processing~\cite{mirzoyan2020dynamic, xu2022strong}. 
 
 
 The applied strain changes the magnetic states, band topologies, and band gaps of carbide and nitride MXenes \cite{shah2022interplay, bafekry2020strain}, whereas transition metal (TM) doping enhances their stability \cite{fatima2020nb}. The TM doped  Cr$_2$C with -F, -O, and -OH functionalization shows the change in electronic and magnetic properties \cite{sun2020tunable}. However, the influence of electron-electron correlation due to the TM doping in these MXenes are still illusive. In addition, the electronic structure with new credible functionalization (-Cl and -H) along with the TM doping is missing. 

Here, we performed DFT calculations of Cr$_2$AlC derived bare, functionalized, strained, and $3d$-electron doped MXenes.  The functionalization makes Cr$_2$C MXene stable, exhibiting magnetic semiconducting behavior. Cr$_2$CF$_2$, a wide band gap semiconductor, manifests super exchange interactions facilitated by F atoms within the layer. Strong spin-phonon coupling found in Cr$_2$CH$_2$ is supported by the distorted Cr spin density that is created by hydrogen environment. Two magnon branches associated to the two different sublattice spins are identified in the ferromagnetic Cr$_2$CO$_2$ and antiferromagnetic Cr$_2$CF$_2$. In addition, the applied tensile strain transforms it from an indirect to a direct band gap semiconductor. Fe or V doped Cr$_2$C remains metallic with FIM ground state, while Fe and V doped Cr$_2$CCl$_2$ and Fe doped Cr$_2$CF$_2$ become a bipolar  semiconductors. In contrast, Cr$_2$CH$_2$ and Cr$_2$C(OH)$_2$ show metallic FIM, whereas Cr$_2$CO$_2$ show half-metallic (metallic) FM with Fe (V) dopants. 	
Importantly, the  underestimated band feature and band magnetism under standard DFT are recovered by correcting self interaction error of 3$d$ electrons by hybrid functional calculations.

First principles calculations were performed using Vienna \textit{Ab initio} Simulation Package (VASP)~\cite{furthmuller1996dimer}  within the projector augmented wave pseudo potential methods. Here, Perdew-Burke-Ernzerhof (PBE) functional~\cite{perdew1996generalized}) was used within the generalized gradient approximation (GGA)~\cite{grimme2006semiempirical}. We used an optimized total  plane wave cut off energy of 500~eV, and 12 $\times$ 12 $\times$ 1 and 25 $\times$ 25 $\times$ 1 \textbf{k}-mesh for the ground state electron density and density of states (DOS) calculations, respectively. A vacuum space of $\approx$ 20 ~{\AA} along the \textit{c}-axis was set. The fully relativistic pseudo-potentials were used for the non-collinear calculations. The GGA with onsite electron correlation (\textit{U}) and exchange (\textit{J}), and standard hybrid functional (HSE06)      \cite {heyd2003hybrid} calculations were performed. We used $U_{eff}$ = $U$ - $J$ = 3~eV \cite{wang2016theoretical}, 3.4~eV \cite{stahl2020critical}, and 4.5~eV \cite{bhandari2021quantum} for Cr, V, and Fe, respectively. The dynamical stability was tested using density functional perturbation theory on the irreducible set of 6 $\times$ 6 $\times$  1 \textbf{q}-mesh.  	
 		
\begin{figure}[!ht]
\centering	
\includegraphics[width=0.47\textwidth]{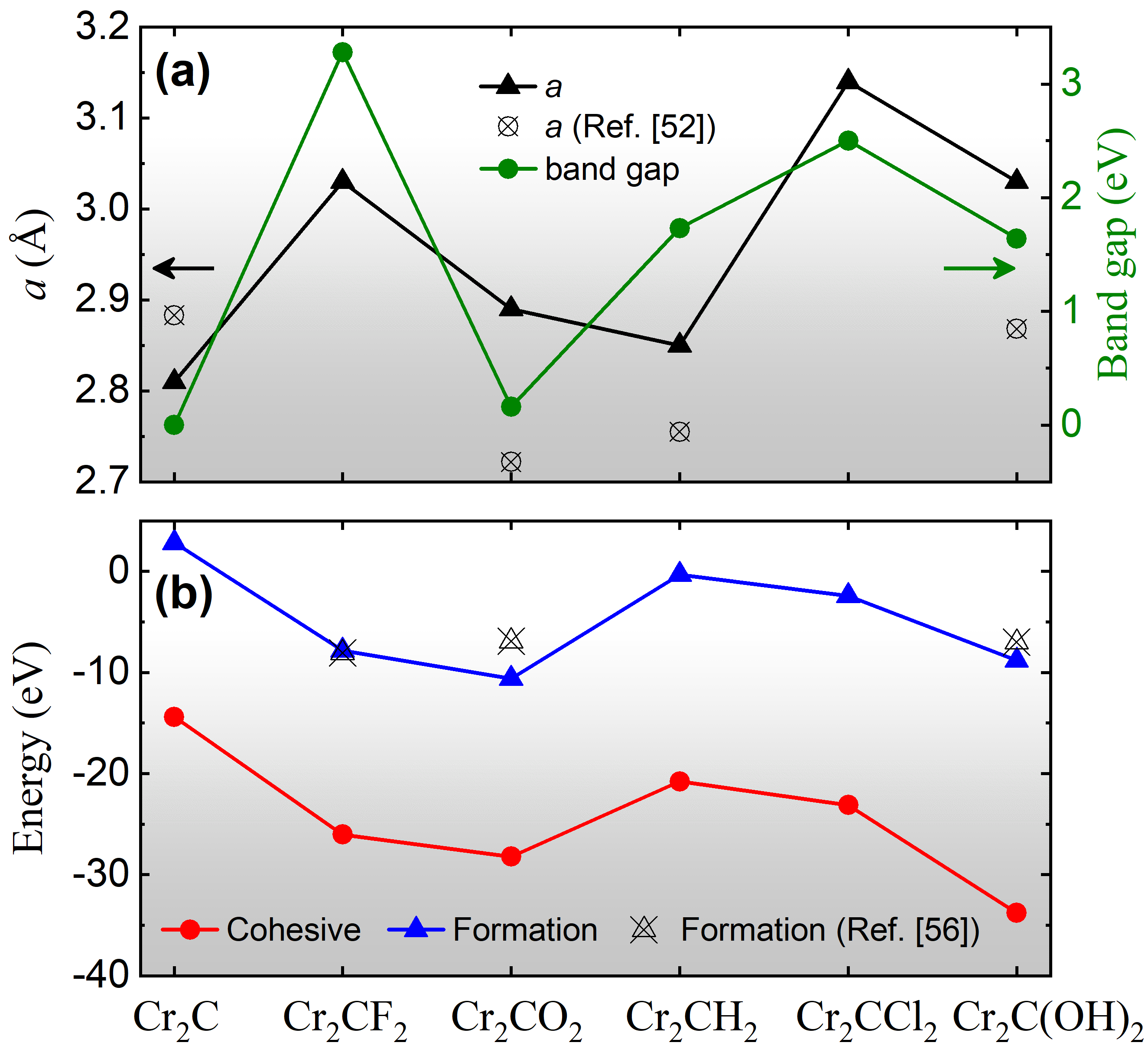}\hfill	
\caption[dop]{(a)-(b) Optimized lattice parameter (black), band gap (green), cohesive energy (blue), and formation energy (red) of the bare and functionalized Cr$_2$C. The literature values are indicated by open symbols.}
\label{figure1}
\end{figure}

We first elucidate the stability of bare, functionalized, and doped Cr$_2$C. The geometry optimization predicts that functional groups prefer to be attached alternatively up and down to the Cr atoms~\cite{wang2017first} in Cr$_2$C. The calculated lattice parameters are in good agreement with previous reports~\cite{schneider2004ab,cui2012first,wang2020uranium, wang2020uranium, wang2020uranium} {(Fig.~\ref{figure1}(a) and Table S1~\cite{supplemental})}.  To study chemical and structural stability, formation and cohesive energies are calculated by using~\cite{limbu2022electronic}

\begin{equation*}
\begin{split}
E_{f/c}(Cr_2CT_2) = {E_\text{tot}(Cr_2CT_2)-2E_\text{bulk/{iso}} (Cr)}\\-{E_\text{bulk/{iso}}(C)} -{2E_\text{bulk/{iso}}(T)},
\end{split}
\end{equation*}

where $E_\text{tot}$ and $E_\text{bulk}$, and $E_\text{iso}$ represent the total energy of optimized systems, and the total energy of isolated atoms.  The cohesive energy of bare Cr$_2$C is negative, however, its formation energy is small but positive (Fig.~\ref{figure1}(a)) indicating that it may require some extra energy such as laser ablation for its synthesis~\cite{akinola2021synthesis, ma2016synthesis}. The negative values of cohesive and formation energies in Cr$_2$C with functional groups (-Cl, -F, -H, -O, and -OH) confirm their structural and chemical stability (Fig.~\ref{figure1}(b)), consistent with a previous study~\cite{khazaei2013novel}.
 
\begin{figure}[!ht]
\centering	
\includegraphics[width=0.47\textwidth]{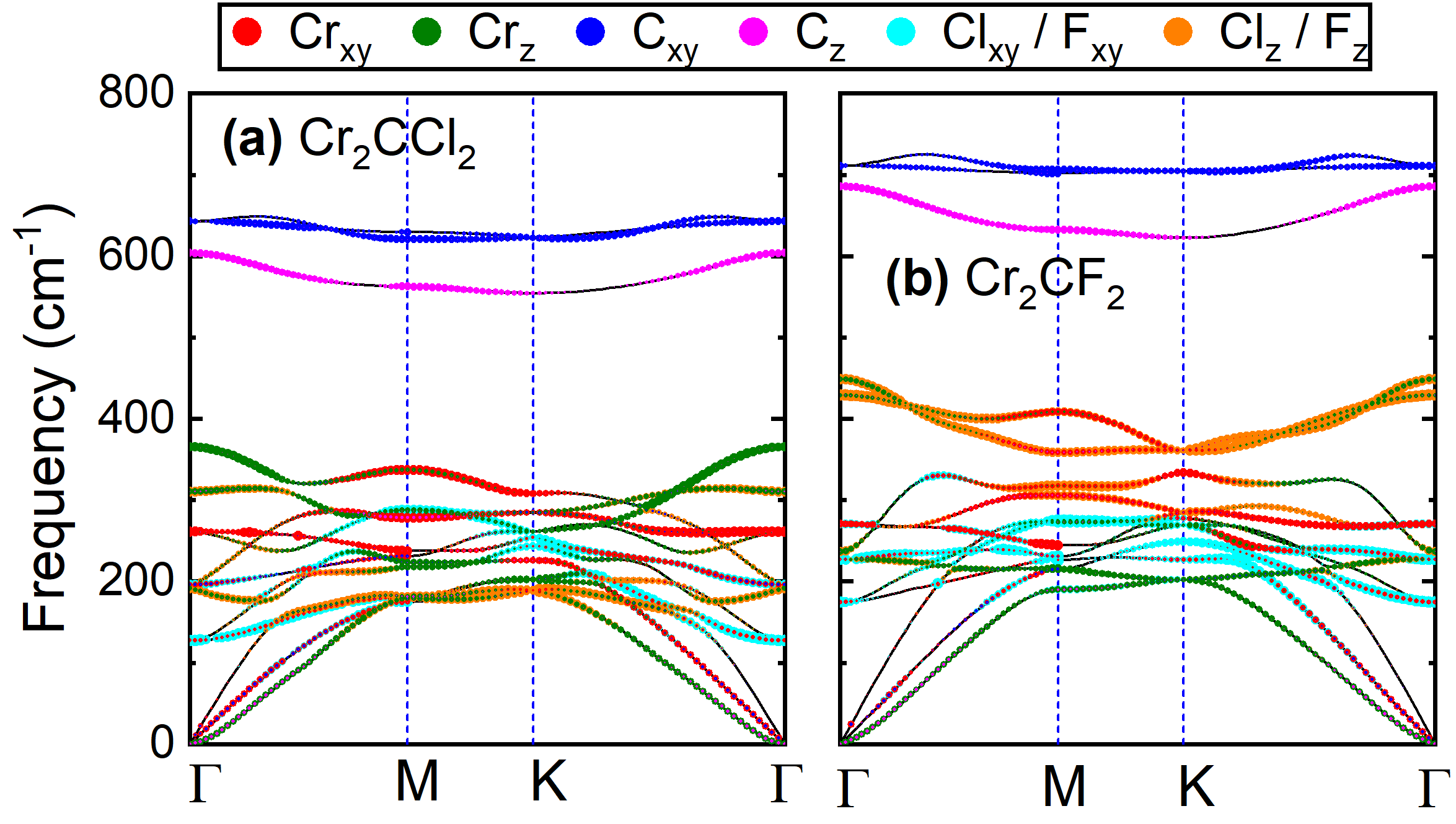}\hfill	
\caption[dop]{Calculated phonon dispersion curves of (a) Cr$_2$CCl$_2$ and (b) Cr$_2$CF$_2$. Different color dots (scaled to the radius) on the phonon spectra signify the contribution from in-plane ($xy$) and out-of-plane ($z$) vibrations of the atoms. The positive phonon frequencies confirm the dynamical stability.}
\label{figure2}
\end{figure}

To check mechanical stability of bare and functionalized, and Fe and V doped Cr$_2$C, we calculate elastic constants using fully relaxed bi-axial strain. The calculated elastic stiffness coefficients {\color{blue}(Table S2 and S4~\cite{supplemental})} fulfill the mechanical stability criteria ($\lvert C_{11}\rvert$ $>$ $\lvert{C_{12}\rvert}$, C$_{11}$ $>$ 0, and C$_{44}$ $>$ 0)~\cite{sharma2022strain}. Calculated phonon frequencies with no imaginary modes confirm the dynamical stability of bare and functionalized Cr$_2$C 
in their respective magnetic ground states 
(Fig.~\ref{figure2} 
and {\color{blue} Fig.~S2}~\cite{supplemental}). 
For example, Cr$_2$CCl$_2$ and Cr$_2$CF$_2$ (Fig.~\ref{figure2}) have similar phonon spectra in low 
and relatively high 
frequency regions. The high frequency modes correspond to vibration of C atoms, while the low frequency modes come from the vibration of other atoms. The vibrations of C atoms in the in-plane and out-of-plane direction are distinctly separated, whereas the vibrations of other atoms are intertwined resulting mixed vibrations in 0-400 cm$^{-1}$ range (Fig.~\ref{figure2}). It is interesting to note that Cr$_2$CH$_2$ shows positive phonon frequencies in both FM and AFM spin configurations, suggesting their dynamical stability ({\color{blue} Fig.~S3(a)-(b)}~\cite{supplemental}). There are two doubly degenerate $E_g$ (in-plane vibrations of Cr and H atoms) and two $A_{1g}$ (out-of-plane vibrations of H atoms) Raman active modes with slight shift in the phonon frequency. The shift is due to the spin lattice interaction, supported from the distorted spin density due to hydrogen, is $\Delta \omega = \lambda \langle \textbf{S}_i.\textbf{S}_j \rangle$, where $\lambda$ is the spin phonon coupling constant, and $\langle \textbf{S}_i.\textbf{S}_j \rangle$ is spin correlation function. Using the maximum shift in phonon frequency and spin correlation function estimated from the spin magnetic moment of Cr, the calculated value of $\lambda$ is 116.5~cm$^{-1}$, which is higher as compared to the available values in 2D magnetic materials~\cite{kozlenko2021spin, xu2022strong}.

The chemical stability of Fe and V doped bare and functionalized Cr$_2$C is investigated by calculating defect formation energy, $E_f= (E_\text{doped}-E_\text{undoped})-(E_\text{dopant}-E_{\text{Cr}})$~\cite{limbu2022electronic}, 
where $E_\text{doped}$, $E_\text{undoped}$, $E_\text{dopant}$, and $E_{\text{Cr}}$ represent the total energies with Fe or V doped, undoped, isolated dopants (Fe or V), and isolated Cr atoms, respectively. 
The negative values of defect formation energy 
indicate that the doped MXenes are chemically stable in their respective magnetic ground states (discussed later) {\color{blue}(Table S3~\cite{supplemental})}. The small positive values (with GGA) in Fe doped  Cr$_2$CF$_2$, Cr$_2$C(OH)$_2$, and Cr$_2$CCl$_2$ reveal the requirement of more energy to stabilize  as  compared to that with V doping.  On the other hand, GGA + U, with U$_\text{eff}$ = 3~eV for Cr and 4.5~eV for Fe, makes these Fe-doped systems chemically stable, indicating the need of onsite electron correlation effect on 3$d$ electrons. 

\begin{figure}[!ht]
\centering			
\includegraphics[width=0.49\textwidth]{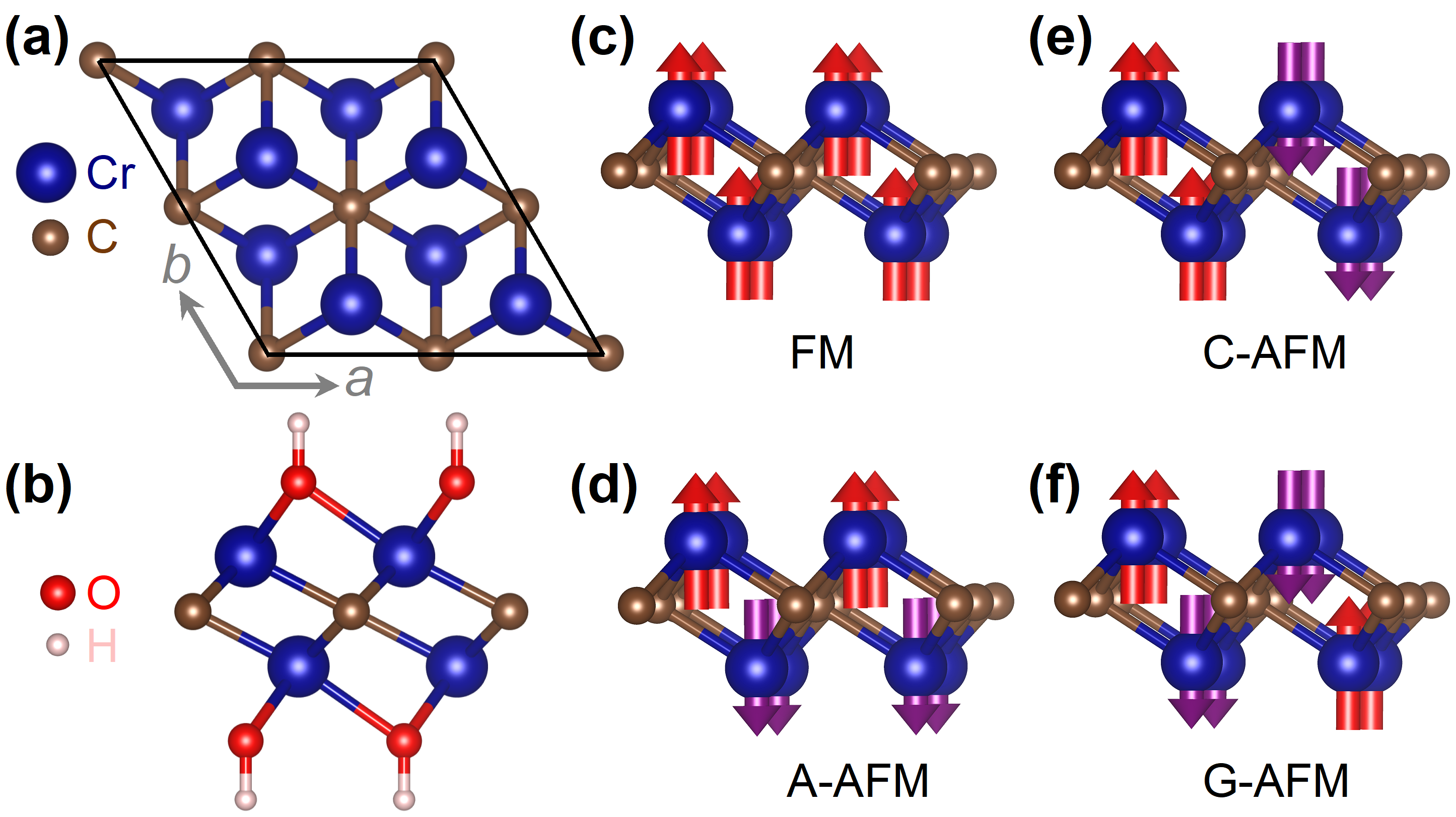}\hfil
\caption[dop]{Bare (a) with top view and -OH functionalized with side view of Cr$_2$C with 2 $\times$ 2 $\times$ 1 super-cell. (c)-(f) Four different possible magnetic configurations of Cr in Cr$_2$C, where arrows on Cr atoms show the direction of spin.}
\label{figure3}
\end{figure}

We now identify the ground state configuration and turn our discussion to electronic properties. The MAX phase (Cr$_2$AlC) is energetically favorable within non magnetic (NM) state under GGA, however, GGA + U makes it AFM as also claimed by Ref.~\cite{dahlqvist2015critical}. Figure~\ref{figure3} shows crystal structures of (a) bare, (b) -OH functionalized, and (c)-(f) different magnetic configurations of Cr$_2$C. The removal of Al from the parent Cr$_2$AlC MAX phase results to the bare Cr$_2$C MXene, which is interlayer FM within the unit cell. However, it becomes intra-layer AFM (C-AFM) with 2 $\times$  2 $\times$ 1 supercell.  -Cl, -F, or -OH functionalization on Cr$_2$C  favors for interlayer AFM (A-AFM), whereas, the -H or -O functionalization  favors for FM within the unit cell.  Interestingly, the -H termination converts it to C-AFM with 2 $\times$ 2 $\times$ 1 supercell. This analysis indicates that the magnetism in Cr$_2$C based MXenes is size dependent, similar in other 2D materials \cite{ABDULHALIM2023170573}.
Both Fe and V doped Cr$_2$C and all functionalized systems are FIM, except the Fe doped Cr$_2$C, and Fe and V doped Cr$_2$CO$_2$, which are FM in the GGA level~{(Table S3~\cite{supplemental})}. However, GGA + U makes all the Fe and V doped systems FIM.

\par Cr$_2$AlC and bare Cr$_2$C show metallicity, while upon -Cl, -F, and -OH functionalization, Cr$_2$C changes from metallic to semi-conducting with indirect band gaps of 0.79~eV, 1.13~eV, and 0.38~eV, respectively {(Fig. S4~\cite{supplemental})}. We note that the onsite electron correlation using GGA + U changes the metallic Cr$_2$CH$_2$ to indirect semiconductor with band gap of 0.84~eV {(Fig. S5~\cite{supplemental})}. There is a formation of a somewhat flat single highest occupied band in Cr$_2$CF$_2$ and Cr$_2$C(OH)$_2$ due to the hybridization of  $d_{z^2}$, $d_{xy}$, and $d_{x^2-y^2}$ states. The SOC splits bands in all cases with minimal changes in the band gap {(Fig. S6~\cite{supplemental})}. HSE06 on the top of GGA significantly increases the band gaps to 2.50~eV, 3.28~eV, and 1.64~eV, for -Cl, -F, and -OH functionalization, respectively. It also changes the metallic and half-metallic nature of -H and -O functionalization to semicondunctor with an indirect band gap of 1.72~eV and 0.16~eV {\color{blue}(Fig. S7~\cite{supplemental})}. 

\begin{figure}[!ht]
\centering
\includegraphics[width=0.47\textwidth]{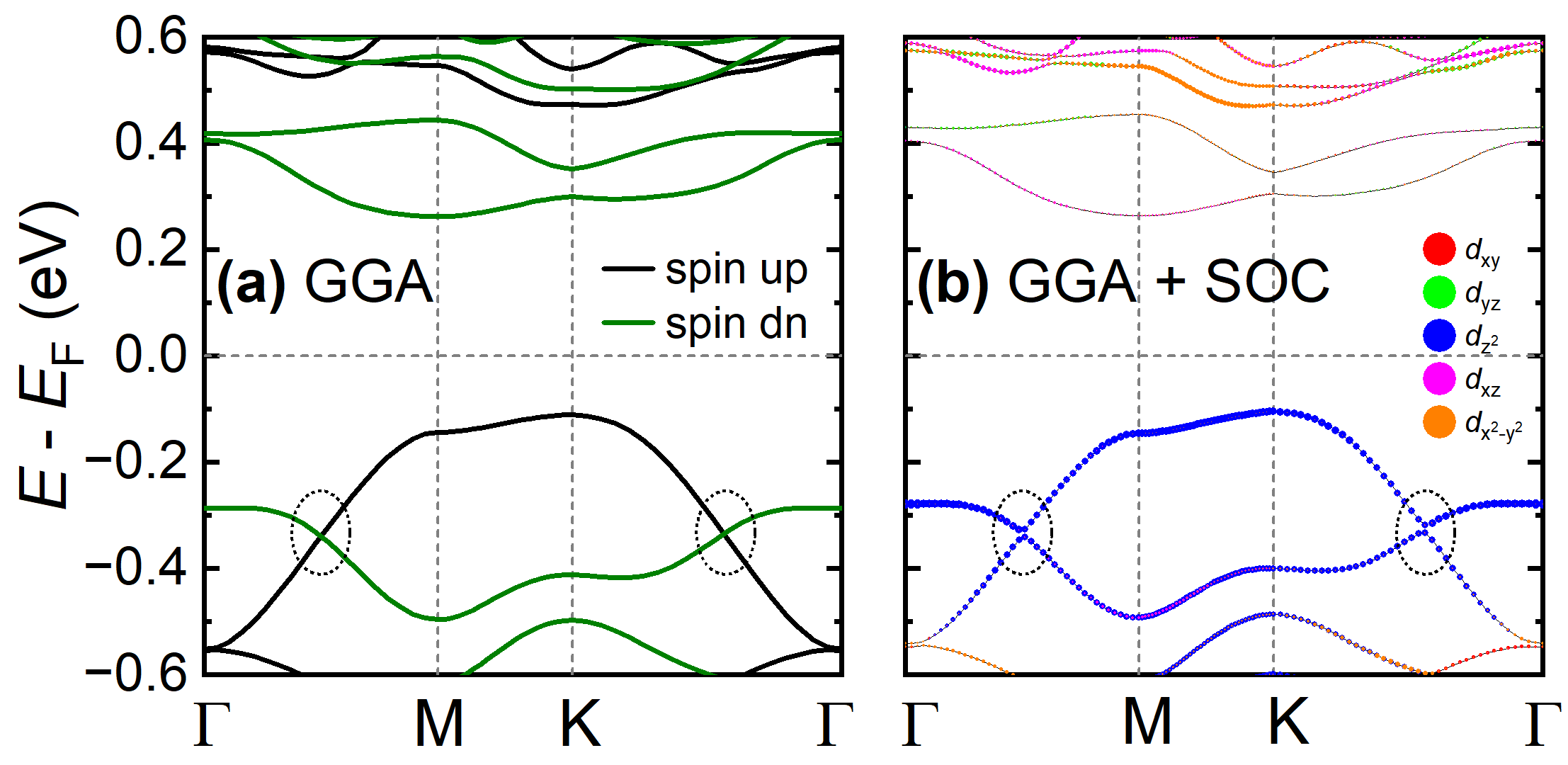}
\caption[dop]{Calculated electronic band structure with (a) GGA and (b) GGA + SOC of Fe doped Cr$_2$CCl$_2$. The spin up and spin down bands cross at two points in (a), whereas the SOC opens a gap in these crossings (b), exhibiting Dirac topology via band inversion.}
\label{figure4}
\end{figure}

Fe and V doped bare Cr$_2$C MXenes are metallic {(Table~S5 and Figs.~S8-S9~\cite{supplemental})}.  In the GGA level, the chlorinated Cr$_2$C, which is an indirect band gap magnetic semiconductor (0.79~eV) changes to a bipolar magnetic semiconductor (0.37~eV) with Fe doping. While the V doped Cr$_2$CCl$_2$ preserves the indirect band gap semi-conducting behavior with reduced gap of 0.18~eV. Here, a flat band appears just below and above (M - K direction) the Fermi level, indicating a zero group velocity and an infinite effective mass. HSE06 increases the band gap of Fe and V doped Cr$_2$CCl$_2$ to 1.80~eV and 1.64~eV, exhibiting bipolar nature {(Fig.~S10~\cite{supplemental})}. In Fe doped case, the SOC admixes spin up and spin down 3$d_{z^2}$ states, opening overall Dirac gaps below the Fermi level along the $\Gamma$ - M and K - $\Gamma$ directions, resulting a band inversion between the two highest valence bands {(Fig.~\ref{figure4})}. By tuning the system, one can shift the Fermi level within the Dirac gaps, thereby making it a quantum anomalous Hall insulator~\cite{doi:10.1126/science.1187485, PhysRevB.98.161111}. The top of the valence and bottom of the conduction bands are contributed by 3$d$ states of Cr and Fe. In the occupied region, the Fe-3$d$ states  are mostly localized in between -6~eV to -8~eV, while Cr-3$d$ states are extended from -8~eV to the Fermi level. 

\par In the GGA level, the fluorinated Cr$_2$C transforms to a direct band gap  semiconductor (0.46~eV at K) with Fe and half-metallic (HM) with V doping~{(Table S5 and Figs.~S8-S9~\cite{supplemental})}. However, the GGA + U makes it an indirect band gap magnetic bipolar semiconductor (1.63~eV) with Fe  and a direct band gap semiconductor (1.48~eV at K) with V doping~{(Table S5 and Figs.~S11-S12~\cite{supplemental})}, while HSE06 increases these band gaps to 2.96~eV and 2.22~eV. Interestingly, in the GGA level, the hydrogenated Cr$_2$C remains metallic with Fe and V doping. However, GGA + U changes it to an indirect band gap semiconductor (0.29~eV at $\Gamma$ - M) with Fe and a direct band gap bipolar semiconductor (0.28~eV at $\Gamma$) with V doping. 
 
The oxygenated Cr$_2$C shows half-metallic/metallic character with Fe/V doping in both the GGA and GGA + U calculations~{(Table~S5, Figs.~S8-S9, and Figs.~S11-S12~\cite{supplemental})}. In Fe-doped case, a metallic character is solely due to the 3$d$ electrons of Cr, whereas the semi-conducting (0.78~eV) band edges are contributed mainly by the hybridized 3$d$ states of Fe and Cr.  Further,  with GGA, Cr$_2$C with -OH functionalization  reduces the indirect band gap from  0.38~eV to 0.04~eV  with Fe doping  and HM with V doping. Here, the GGA + U and HSE06 calculations show semi-conducting behavior with significant band gaps {\color{blue}(Figs.~S6-S12~\cite{supplemental})}. 

\begin{figure}[!ht]
\centering	
\includegraphics[width=0.48\textwidth]{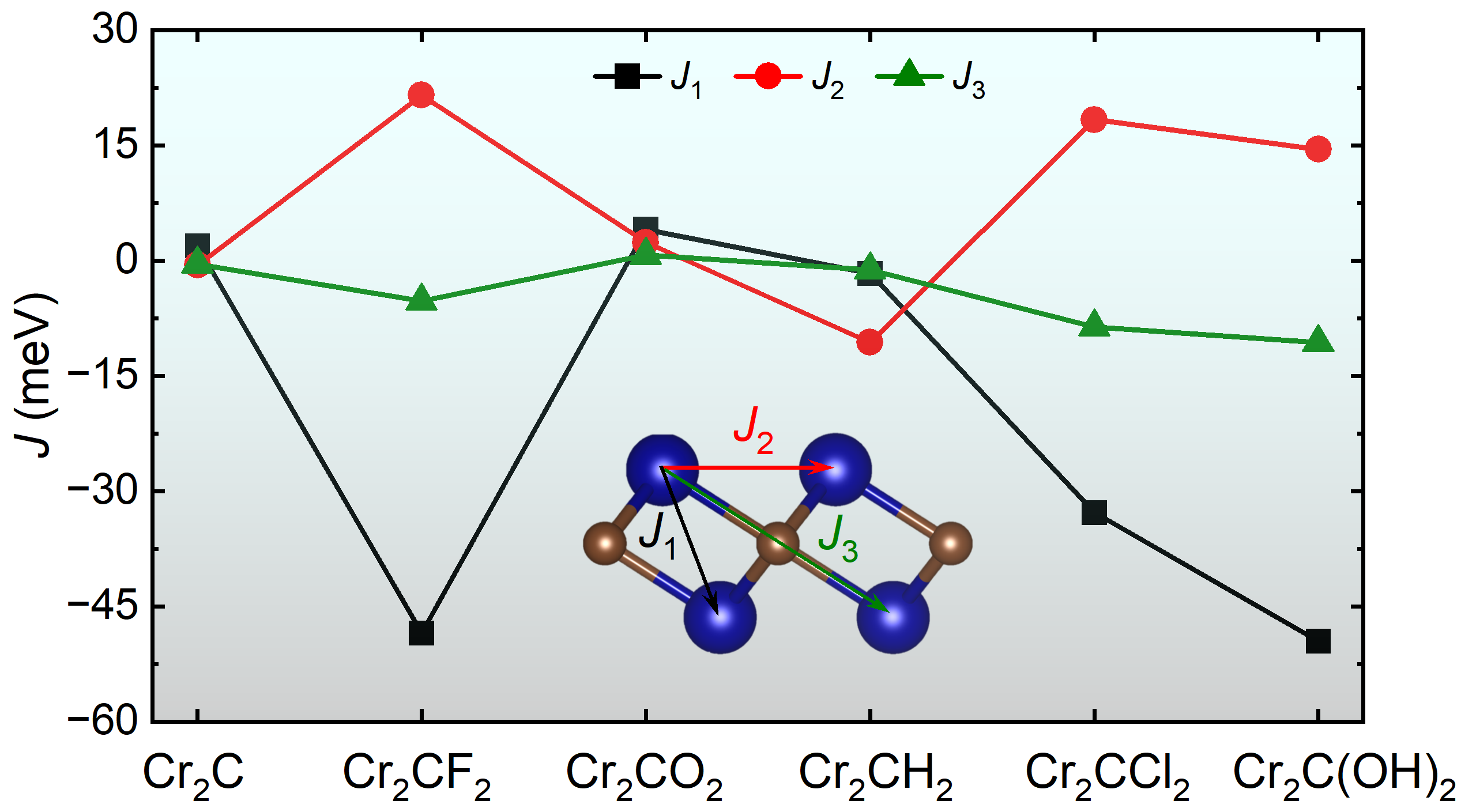}
\caption[dop]{Exchange interactions ($J_1$, $J_2$, and $J_3$) of the bare and functionalized Cr$_2$C MXenes. The black, red, and green lines represent the trend of  $J_1$ (inter-layer), $J_2$ (intra-layer), and $J_3$ (inter-layer).}
\label{figure5}
\end{figure}

We now analyse the magnetic properties. With GGA, Cr$_2$C is FM with total magnetic moment (MM) of 1 $\mu_B$/f.u., while HSE06 increases it to 1.5 $\mu_B$/f.u. Interestingly, increasing the unit cell size to 2 $\times$ 2 $\times$ 1, the FM alignment changes to FIM ground state with an unequal and oppositely aligned intra-layer Cr moments (1.22~$\mu_B$/atom and -0.37~$\mu_B$/atom), as also conformed by the magnetic exchange interactions (Fig.~\ref{figure5}) discussed below. 

The exchange interactions $J_1$ (inter-layer), $J_2$ (intra-layer), and $J_3$ (inter-layer) are calculated by employing the Heisenberg spin Hamiltonian.
\begin{equation}
H_\text{spin} = -\sum_{i\neq j}J_1 \mathbf{S}_i \cdot\mathbf{S}_j - \sum_{k\neq l}J_2 \mathbf{S}_k \cdot\mathbf{S}_l -  \sum_{m\neq n}J_3 \mathbf{S}_m \cdot\mathbf{S}_n,
\label{Hamiltonian}
\end{equation}
where ($i$, $j$), ($k$, $l$), and ($m$, $n$) represent first-, second-, and third-nearest-neighbor Cr atoms, respectively. The total energies of different magnetic configurations {\color{blue}(Fig. \ref{figure3}(c)-(f))} are then used to the Heisenberg model in the following expressions  \cite{lv2020monolayer}.
\vspace{-0.3cm}
\begin{equation*}
E_{\text{FM}/\text{A-AFM}} = E_{0} -  (\pm3 J_1 +6 J_2 \pm3 J_3)|S^2|
\end{equation*}
\vspace{-0.8cm}
\begin{equation*}
E_{\text{C-AFM}/\text{G-AFM}} = E_{0} - (\pm J_1 -2J_2 \mp 3J_3)|S^2|
\end{equation*}
where, $S = \frac{3}{2}$ for Cr$^{3+}$ ion from Hund's rule. 
Solving above equations, we get
\vspace{-0.1cm}
\begin{equation*}
J_1 = \frac{E_\text{A-AFM}+ E_\text{G-AFM} - E_\text{FM}- E_\text{C-AFM}}{8|S|^2}
\end{equation*}
\vspace{-0.5cm}
\begin{equation*}
J_2 = \frac{E_\text{C-AFM}+ E_\text{G-AFM} - E_\text{FM}- E_\text{A-AFM}}{16|S|^2}
\end{equation*}
\vspace{-0.5cm}
\begin{equation*}
J_3 = \frac{3E_\text{C-AFM} + E_\text{A-AFM} -  3E_\text{G-AFM}-E_\text{FM}}{24|S|^2}.
\end{equation*}

The calculated values of exchange interactions for the bare Cr$_2$C follow the trend of $J_1 > 0$, $J_2 < 0$, and $J_3 < 0$ (Fig.~\ref{figure5}) and show both parallel and antiparallel interlayer as well as antiparallel intra-layer Cr moment alignments, confirming the FIM ground state. The Cl functionalized Cr$_2$C is A-AFM with 2.21~$\mu_B$/Cr, which increases to 2.82~$\mu_B$/Cr with HSE06~{(Table~S6~\cite{supplemental})}. In this case the exchange interactions follow the trend of $J_1 < 0$, $J_2 > 0$, and $J_3 < 0$, indicating strong interlayer AFM alignments and intra-layer FM alignments between Cr moments~{(Table~S7~\cite{supplemental})}. 
						
The fluorinated Cr$_2$C is also A-AFM with 2.54~$\mu_B$/Cr, which increases to 2.97~$\mu_B$/Cr with HSE06 {(Table~S6~\cite{supplemental})}. The calculations show significantly high value of energy difference between A-AFM and NM configurations resulting in large magnetic exchange interactions {(Table~S6~\cite{supplemental})} as also indicated in Ref.~\cite{kumar2017tunable}. These exchange interactions follow the trend of $J_1 < 0$, $J_2 > 0$, and $J_3 < 0$ with strong antiparallel inter-layer and relatively weak parallel intra-layer Cr moment alignments. Unexpectedly, Cr$_2$CF$_2$ shows super exchange interactions between intra-layer and direct exchange interactions between inter-layer Cr moments. The 3$d$ states of Cr$^{3+}$ anion hybridize with 2$p$ states of F$^{1-}$ cation exhibiting ferromagnetic super exchange interactions~\cite{jiang2021recent}. In this case, the anion and cation are connected by $\sim$~90$^\circ$ with relatively large Cr-Cr distance (3.04~\AA) and small but non negligible induced magnetic moment in F$^{1-}$ cation as suggested  by Goodenough-Kanamori-Anderson~\cite{anderson1959new,goodenough1955theory, kanamori1960crystal}. The  angle between intra-layer cation (Cr$^{3+}$) through C$^{4-}$ anion is larger than 90$^\circ$ (97.06$^\circ$), which helps decrease relative angle between carbon connected inter-layer Cr atoms reducing Cr-Cr distance (2.68~\AA) and enabling direct exchange interactions with AFM spin alignments.

\par  The Cr$_2$CO$_2$ is FM with 2.09 $\mu_B$/Cr (2.43~$\mu_B$/Cr) with GGA (HSE06) {(Table~S6~\cite{supplemental})}. The angle between interlayer Cr atoms through C is $\sim$ 90$^{\circ}$, exhibiting  FM super exchange interaction via the induced MM~(-0.20~$\mu_B$) of C. Here, the intra-layer distance between Cr atoms (2.89~\AA) is shorter than that of inter-layer distance (3.04~\AA), confirming the direct intra-layer FM exchange interaction. In this case, the exchange interactions are all positive (FM) with $J_1 > J_2 > J_3$ {(Fig.~\ref{figure5} and Table~S7~\cite{supplemental})}. On the other hand, Cr$_2$C(OH)$_2$ is A-AFM with MM of 2.40~$\mu_B$/Cr (2.89~$\mu_B$/Cr) with GGA (HSE06), and Cr$_2$CH$_2$ is FM with total MM of 2.97~$\mu_B$/f.u. with GGA.  In Cr$_2$C(OH)$_2$, the calculated exchange interactions are higher in magnitude than that with other functional groups. Interestingly, while in-plane doubling of the unit cell, the FM alignment changes to intra-layer FIM in Cr$_2$CH$_2$ with GGA. 

\begin{figure}[!ht]
\centering	
\includegraphics[width=0.47\textwidth]{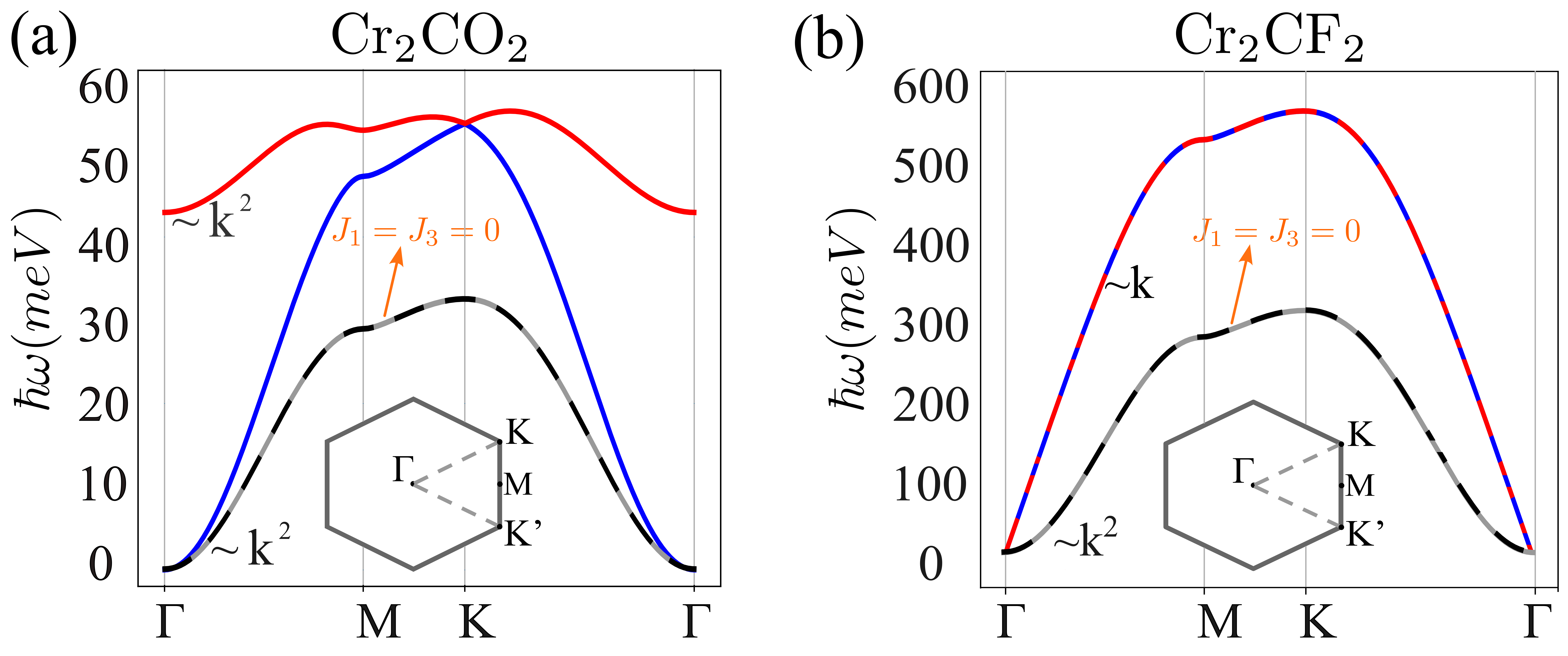}\hfill
\caption[dop]{Magnon dispersion for the ferromagnetic case Cr$_2$CO$_2$ (a) and antiferromagnetic Cr$_{2}$CF$_{2}$ (b). The black and gray dashed curves correspond to $J_1 = J_3 = 0$~{(Fig.~S1 \cite{supplemental})}.}
\label{fig6}
\end{figure}

Using the exchange interactions obtained from the calculations described above, we determine the magnon energy dispersion via the application of the Holstein-Primakoff transformation~\cite{PhysRev.58.1098, prabhakar2009spin} within Hamiltonian (Eq.~\ref{Hamiltonian})~\cite{supplemental} for the ferromagnetic (FM) Cr$_{2}$CO$_{2}$ and antiferromagnetic (A-AFM) Cr$_{2}$CF$_{2}$ (Fig.~\ref{fig6}a and ~\ref{fig6}b). Two magnon branches are identified, which are associated to two different sublattice spins predominantly by $J_{1}$ and $J_{2}$. The degeneracy at the $\mathbf{K}$ ($\mathbf{K'}$) points in the FM phase are not lifted as the Dzyaloshinskii-Moriya interactions (DMI) are not included, however the two branches in the A-AFM state are fully degenerate without the DMI. The large magnon energies arise from the large exchange interaction values for the FM and A-AFM ~{(Table S7~\cite{supplemental})} combined with the large number of neighbors, $Z_{1}$ = $Z_{3}$ = 3 and $Z_{2}$ = 6, and $S$ = 3/2. Two uncoupled FM (black and grey curves) dispersions (Fig.~\ref{fig6}) are found while setting $J_{1}$ = $J_{3}$ = 0 ~{(Fig.~S1 \cite{supplemental})}. The FM and AFM display quadratic and linear dispersions, respectively, around $\Gamma$ (Fig.~\ref{fig6}).

With GGA, the Fe doped Cr$_2$C is FM  with total MM of 5.68~$\mu_B$/cell (1.83~$\mu_B$/Fe), while the V doped Cr$_2$C remains FIM {(Table~S8~\cite{supplemental})}. GGA + U makes the Fe doped system to FIM within the layer along with large Cr moment and small but non-negligible induced C moment. The individual MMs of each Cr atom are larger in V doped than that of Fe doped. With GGA, Fe and V doped Cr$_2$CCl$_2$, 
Cr$_2$CF$_2$, and Cr$_2$C(OH)$_2$ exhibit FIM characteristic in which Cr and the dopants align in intra-layer FM configuration. The MM of each Cr atom is relatively larger in Fe doped as compared to V doped Cr$_2$CCl$_2$ and Cr$_2$CF$_2$, as expected from the Slater-Pauling relation~\cite{williams1983generalized}. In contrast, Fe/V doped Cr$_2$CH$_2$ is FIM within the layer, while Cr$_2$CO$_2$ displays FM character.

\par We now explain the influence of compressive (up to -10\%) and tensile (up to +10\%) strain in the bare and functionalized Cr$_2$C. With GGA, the bare Cr$_2$C remains magnetic metal up to 4\% compressive strain and 10\% tensile strain, while it becomes NM from 4\% to 10\% compressive strain~{(Fig.~S13 and Table~S9 \cite{supplemental})}. On increasing the tensile strain up to 4\% with 2\% increment, its total MM decreases from 1.02~$\mu_B$ to 0.97~$\mu_B$, and 0.87~$\mu_B$. However, further increase in tensile strain from 4\% to 10\% with the same rate, the MM increases non-linearly to 0.73~$\mu_B$, 0.80~$\mu_B$, and 1.41~$\mu_B$.  

With the increase of compressive strain, the band gap of Cr$_2$C(OH)$_2$ diminishes and becomes semi-metallic at 8\%, and finally turns into a metal at 10\%. Under the tensile strain, the band gap increases preserving its AFM character {(Fig.~S14                  \cite{supplemental})}. In Cr$_2$CF$_2$, the compressive strain continuously decreases the band gap, preserving its indirect nature, whereas a 2\% tensile strain transforms it to a direct bandgap with a non-dispersive band just below the Fermi level~{(Fig.~S15~\cite{supplemental})}. Further increasing the tensile strain, the band gap retrieves back to the indirect type with distorted flat bands. On the other hand, in Cr$_2$CCl$_2$, compressive strain (at 2\%) first increases and then (at 4\%) decreases  the band gap~{(Fig.~S16~\cite{supplemental})}. Further increase in the compressive strain, we find a continuous decrease in the band gap, which turns it from an AFM semiconductor to a FM metal at 10\% {(Table~S9~\cite{supplemental})}, whereas there is a continuous decreases in the indirect band gap with increased tensile strain.

HSE06 calculations show that semiconducting Cr$_2$CO$_2$ is sensitive to the compressive strain as compared to the tensile strain~{(Fig.~S17~\cite{supplemental})}. Interestingly, 2\% compressive strain  increases the band gap and transforms it to a bipolar magnetic semiconductor with an indirect band gap (at $\Gamma$ and K), while 4\%, 8\%, and 10\% compressive strains make it to a semi-metal, indirect band gap bipolar semiconductor, and indirect band gap semiconductor, respectively. Whereas, a 2\% and higher tensile strain changes the semi-conducting Cr$_2$CO$_2$ to the half-metallic. On the other hand, a 2\% compressive strain decreases the band gap of Cr$_2$CH$_2$~{(Fig.~S18~\cite{supplemental})}. Interestingly, it becomes metallic at 4\%, semiconducting at 6\%, and again metallic beyond 8\%. In contrast to compressive strain, a 2\% tensile strain changes the indirect band gap to a direct band gap at the $\Gamma$ point. Further increasing the tensile strain, the band gap increases at 4\% and continuously decreases beyond 6\%.
 
In conclusion, we report a comprehensive study of structural, electronic, and magnetic properties of bare, functionalized, strained, and Fe or V doped Cr$_2$C MXenes from first principles.  The chemically unstable bare Cr$_2$C becomes stable on functionalization and doping, and the negative cohesive energies, positive phonon frequencies, and favorable elastic constants confirm their structural, dynamical, and mechanical stabilities. The functionalized  Cr$_2$C changes from  FM metal to AFM semiconductors, except for Cr$_2$CO$_2$, which is a FM semiconductor with split magnon branches associated to sublattice spins. The onsite electron correlation effect within 3$d$-electrons  plays a key role to confirm the magnetic semiconducting behavior, viz. the Cr$_2$CF$_2$ stands out as an indirect wide band gap AFM semiconductor with two AFM magnon branches, exhibiting super exchange interaction mediated by F atoms within the layer. In addition, the applied tensile strain transforms it from an indirect to a direct band gap semiconductor. Strong spin-phonon coupling found in Cr$_2$CH$_2$ is supported by the distorted Cr spin density that is created by hydrogen environment. Fe or V doped Cr$_2$C remains metallic with FIM ground state, while Fe and V doped Cr$_2$CCl$_2$ and Fe doped Cr$_2$CF$_2$ become bipolar semiconductors, respectively. In Fe doped  Cr$_2$CCl$_2$, the SOC admixes spin up and spin down 3$d_{z^2}$ states, opening overall Dirac gaps below the Fermi level along the $\Gamma$ - M and K - $\Gamma$ directions, resulting a band inversion between the two highest valence bands. Cr$_2$CH$_2$ and Cr$_2$C(OH)$_2$ show metallic FIM whereas Cr$_2$CO$_2$ show half-metallic (metallic) FM with Fe (V) dopants. The calculated magnetic exchange interactions ($J_1$, $J_2$, and $J_3$) confirm the magnetic ground states provided by the total energy analysis. The large values of effective exchange interactions also confirm functionalized Cr$_2$C as potential high temperature magnetic semiconductors. 			

The work on magnetic interactions, spin phonon coupling, and magnon dispersions is supported as part of the Center for Energy Efficient Magnonics, an Energy Frontier Research Center funded by the U.S. Department of Energy, Office of Science, Basic Energy Sciences, under Award number DE-AC02-76SF00515. The basic \textit{ab initio} part of the work is supported from the Department of Energy under Grant number DE-SC0016379. We acknowledge use of the computational facilities on the Frontera supercomputer at the Texas Advanced Computing Center via the pathway allocation, DMR23051, and the Argon high-performance computing system at the University of Iowa.
\bibliography{references}
\end{document}


\title{Supplemental Material \\ 
Functionalized Cr$_2$C  MXenes: Novel Magnetic Semiconductors} 
\author{Yogendra Limbu}
\affiliation{Department of Physics and Astronomy, University of Iowa, Iowa City, Iowa 52242, USA}
\author{Hari Paudyal}
\affiliation{Department of Physics and Astronomy, University of Iowa, Iowa City, Iowa 52242, USA}
\author{Eudes Gomes da Silva}
\affiliation{Department of Physics and Astronomy, University of Iowa, Iowa City, Iowa 52242, USA}
\author{Denis R. Candido}
\affiliation{Department of Physics and Astronomy, University of Iowa, Iowa City, Iowa 52242, USA}
\author{Michael E. Flatt\'e}
\affiliation{Department of Physics and Astronomy, University of Iowa, Iowa City, Iowa 52242, USA}
\affiliation{Department of Applied Physics, Eindhoven University of Technology, Eindhoven, The Netherlands}
\author{Durga Paudyal}
\affiliation{Department of Physics and Astronomy, University of Iowa, Iowa City, Iowa 52242, USA}

\date{\today}
\maketitle
\section{Magnon dispersion}
The spin-spin interacting Hamiltonian is given by 

\begin{equation}
{\cal H}_{{\rm spin}}=-J_{1}\sum_{\left\langle i,j\right\rangle }\textbf{S}\left(\mathbf{R}_{i}\right)\cdot\textbf{S}\left(\mathbf{R}_{j}\right)-J_{2}\sum_{\left\langle \left\langle i,j\right\rangle \right\rangle }\textbf{S}\left(\mathbf{R}_{i}\right)\cdot\textbf{S}\left(\mathbf{R}_{j}\right)-J_{3}\sum_{\left\langle \left\langle \left\langle i,j\right\rangle \right\rangle \right\rangle }\textbf{S}\left(\mathbf{R}_{i}\right)\cdot\textbf{S}\left(\mathbf{R}_{j}\right), \label{Hamiltonian}
\end{equation}
where $\left\langle i,j\right\rangle $, $\text{\ensuremath{\left\langle \left\langle i,j\right\rangle \right\rangle }},$and
$\text{\ensuremath{\left\langle \left\langle \left\langle i,j\right\rangle \right\rangle \right\rangle }}$
are the sum over the first, second and third nearest-neighbor, respectively.
We note that we are not considering dipole-dipole interaction among different spins, as all the exchange parameters are at least one order of magnitude larger
than the dipole-dipole interaction. The crystal structure is formed by a hexagonal lattice with two Cr atoms per unit cell (defined by
sublattices $A$ and $B$).

The corresponding Bravais lattice is defined by the unit vectors $\mathbf{a_{1}}=\frac{a}{2}\left(3,\sqrt{3}\right)$ and $\mathbf{a_{2}}=\frac{a}{2}\left(3,-\sqrt{3}\right)$ yielding $\mathbf{R_{i}}=n\mathbf{a_{1}}+m\mathbf{a_{2}}$, n, m $\in\mathtt{\mathbb{Z}}$.
The positions of the Cr$_{A}$ and Cr$_{B}$ atoms are then given
by
\begin{align}
\mathbf{R_{i}^{A}}&=n\mathbf{a_{1}}+m\mathbf{a_{2}},\\ \mathbf{R_{i}^{A}}&=n\mathbf{a_{1}}+m\mathbf{a_{2}}+\frac{\mathbf{a_{1}}+\mathbf{a_{2}}}{3}
\end{align}

The reciprocal lattice wavevectors are given by $\mathbf{K_{1}}=\frac{2\pi}{3a}\left(1,\sqrt{3}\right)$ and $\mathbf{K_{2}}=\frac{2\pi}{3a}\left(1,-\sqrt{3}\right)$. 

\begin{figure*}[!ht]
\centering	
\includegraphics[width=0.7\textwidth]{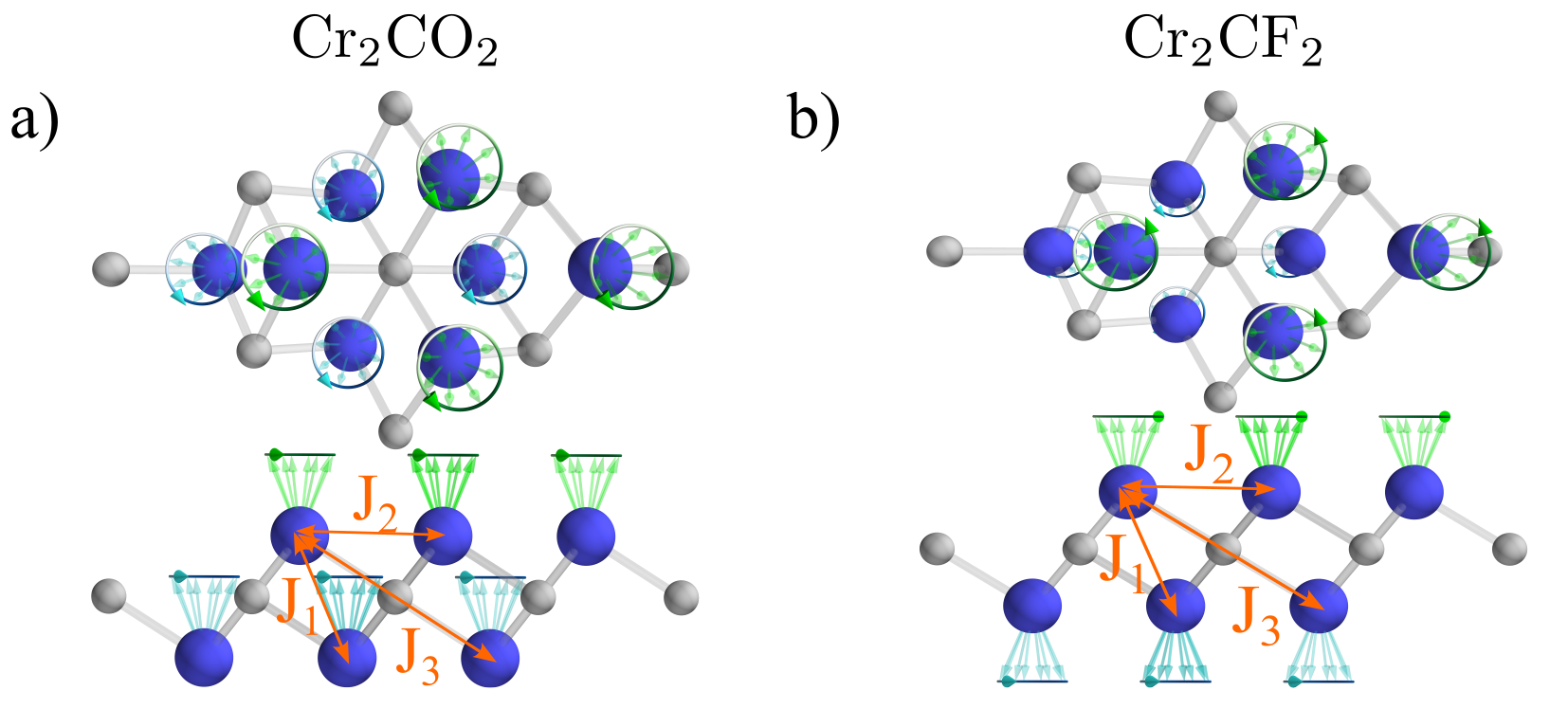}\hfill	
\caption[dop]{Supercell for the ferromagnetic a) (Cr$_{2}$CO$_{2}$) and antiferromagnetic A-AFM (Cr$_{2}$CF$_{2}$). The orange arrows indicate the exchange between first (J$_{1}$), second (J$_{2}$) and third (J$_{3}$) neighbors. The green and blue arrows indicate spin orientation.}
\label{fig-neighbors}
\end{figure*}

\subsection{Ferromagnetic Cr$_{2}$CO$_{2}$}

Here we obtain the magnon dispersion for the ferromagnetic Cr$_{2}$CO$_{2}$, with the ground state
composed of spins all aligned parallel. To obtain the excitation energy for
the magnons, we will use $S_{\pm}^{\mu}\left(\mathbf{R}_{i}^{\mu}\right)=S_{x}^{\mu}\left(\mathbf{R}_{i}^{\mu}\right)\pm iS_{y}^{\mu}\left(\mathbf{R}_{i}^{\mu}\right)$,
and the Hostein-Primakoff approximation [1, 2]

\begin{align}
S_{+}^{\mu}\left(\mathbf{R}_{i}^{\mu}\right)=S_{i,+}^{\mu} & =\sqrt{2S}\left(1-\frac{\mu_{i}^{\dagger}\mu_{i}}{2S}\right)^{1/2}\mu_{i}\approx\sqrt{2S}\mu_{i},\\
S_{-}^{\mu}\left(\mathbf{R}_{i}^{\mu}\right)=S_{i,-}^{\mu} & =\sqrt{2S}\left(1-\frac{\mu_{i}^{\dagger}\mu_{i}}{2S}\right)^{1/2} \mu_{i}^{\dagger} \approx\sqrt{2S}\mu_{i}^{\dagger},\\
S_{z}^{\mu}\left(\mathbf{R}_{i}^{\mu}\right)=S_{i,z}^{\mu} & =S-\mu_{i}^{\dagger}\mu_{i},
\end{align}
with magnon operators $\mu_{i}=a_{i},b_{i}$, obeying $\left[\mu_{i},\mu_{i}^{\dagger}\right]=1$.
To diagonalize the Hamiltonian (Eq.~\ref{Hamiltonian}), we expand our magnon operators as 
\begin{align}
\mu_{i}=\frac{1}{\sqrt{N}}\sum_{\mathbf{k}}e^{i\mathbf{k}\cdot\mathbf{R}_{i}^{\mu}}\mu_{\mathbf{k}}, & \qquad\mu_{i}^{\dagger}=\frac{1}{\sqrt{N}}\sum_{\mathbf{k}}e^{-i\mathbf{k}\cdot\mathbf{R}_{i}^{\mu}}\mu_{\mathbf{k}}^{\dagger},
\end{align}
where $N$ is the total number of spins in each sublattice, and $\mathbf{k}$ is the 2D wavevector. For the ferromagnetic case, we have three nearest (NN) and next-next-next-nearest (NNNN) neighbors  interacting spins from different sub-lattices, i.e., ${\cal Z}_{1}={\cal Z}_{3}=3$.
The vectors connecting them are $\mathbf{T}_{1}=(\mathbf{a_{1}+a_{2}})/3$, $\mathbf{T}_{2}=\left(\mathbf{a_{2}}-2\mathbf{a_{1}}\right)/3$ and $\mathbf{T}_{3}=\left(\mathbf{a_{1}}-2\mathbf{a{2}}\right)/3$ for NN, yielding

\begin{equation}
{\cal H}_{1}=-J_{1}\sum_{\left\langle i,j\right\rangle }\textbf{S}\left(\mathbf{R}_{i}\right)\cdot\textbf{S}\left(\mathbf{R}_{j}\right)\approx-J_{1}S^{2}N{\cal Z}_{1}+{\cal Z}_{1}J_{1}S\sum_{\mathbf{k}}\left(a_{\mathbf{k}}^{\dagger}\quad b_{\mathbf{k}}^{\dagger}\right)\left(\begin{array}{cc}
1 & -\gamma_{\mathbf{k}}^{\left(1\right)}\\
-\gamma_{\mathbf{-k}}^{\left(1\right)} & 1
\end{array}\right)\left(\begin{array}{c}
a_{\mathbf{k}}\\
b_{\mathbf{k}}
\end{array}\right),
\end{equation}
with $\gamma_{\mathbf{k}}^{\left(1\right)}=\frac{1}{{\cal Z}_{1}}\sum_{j}e^{i\mathbf{k}\cdot\mathbf{T}_{j}}$.
Alternatively, for the next nearest neighbor (NNN) Hamiltonian, we
have six interacting spins from the same sublattice, i.e., ${\cal Z}_{2}=6$.
The vectors connecting them are $\mathbf{L}_{1}=\mathbf{a}$, $\mathbf{L}_{2}=-\mathbf{a}$,
$\mathbf{L}_{3}=\mathbf{b}$, $\mathbf{L}_{4}=-\mathbf{b}$, $\mathbf{L}_{5}=\mathbf{a}+\mathbf{b}$,
and $\mathbf{L}_{6}=-\left(\mathbf{a}+\mathbf{b}\right)$. The corresponding
Hamiltonian reads
\begin{align}
{\cal H}_{2}=-J_{2}\sum_{\left\langle \left\langle i,j\right\rangle \right\rangle }\textbf{S}\left(\mathbf{R}_{i}\right)\cdot\textbf{S}\left(\mathbf{R}_{j}\right) & \approx-J_{2}S^{2}N{\cal Z}_{2}\\
 & +{\cal Z}_{2}\frac{J_{2}}{2}S\sum_{\mathbf{k}}\left(a_{\mathbf{k}}^{\dagger}\quad b_{\mathbf{k}}^{\dagger}\right)\left(\begin{array}{cc}
2-\gamma_{\mathbf{-k}}^{\left(2\right)}-\gamma_{\mathbf{k}}^{\left(2\right)} & 0\\
0 & 2-\gamma_{\mathbf{-k}}^{\left(2\right)}-\gamma_{\mathbf{k}}^{\left(2\right)}
\end{array}\right)\left(\begin{array}{c}
a_{\mathbf{k}}\\
b_{\mathbf{k}}
\end{array}\right)\nonumber,
\end{align}
with $\gamma_{\mathbf{k}}^{\left(2\right)}=\frac{1}{{\cal Z}_{2}}\sum_{j}e^{i\mathbf{k}\cdot\mathbf{L}_{j}}$. 
The third neighbors are $\mathbf{P}_{1}=(\mathbf{a_{1}+a_{2}})/3$, $\mathbf{P}_{2}=\left(\mathbf{a_{2}}-2\mathbf{a_{1}}\right)/3$ and $\mathbf{P}_{3}=\left(\mathbf{a_{1}}-2\mathbf{a_{2}}\right)/3$, with Hamiltonian given by

\begin{equation}
{\cal H}_{3}=-J_{3}\sum_{\left\langle \left\langle \left\langle i,j\right\rangle \right\rangle \right\rangle  }\textbf{S}\left(\mathbf{R}_{i}\right)\cdot\textbf{S}\left(\mathbf{R}_{j}\right)\approx-J_{3}S^{2}N{\cal Z}_{3}+{\cal Z}_{3}J_{3}S\sum_{\mathbf{k}}\left(a_{\mathbf{k}}^{\dagger}\quad b_{\mathbf{k}}^{\dagger}\right)\left(\begin{array}{cc}
1 & -\gamma_{\mathbf{k}}^{\left(3\right)}\\
-\gamma_{\mathbf{-k}}^{\left(3\right)} & 1
\end{array}\right)\left(\begin{array}{c}
a_{\mathbf{k}}\\
b_{\mathbf{k}}
\end{array}\right),
\end{equation}
and $\gamma_{\pm \mathbf{k}}^{(3)}=\frac{1}{Z_{2}}\sum_{j}e^{\pm i\mathbf{k}\cdot \mathbf{P}_{j}^{(3)}}$
The final Hamiltonian reads

\begin{equation}
{\cal H}\approx S\sum_{k}\left(a_{\mathbf{k}}^{\dagger}\quad b_{\mathbf{k}}^{\dagger}\right)\left(\begin{array}{cc}
J_{1}Z_{1}+J_{3}Z_{3}-J_{2}Z_{2}(\gamma_{k}^{(2)}+\gamma_{-k}^{(2)}-2) & -J_{1}Z_{1}\gamma_{k}^{(1)}+J_{3}Z_{3}\gamma_{k}^{(3)}\\
-J_{1}Z_{1}\gamma_{-k}^{(1)}+J_{3}Z_{3}\gamma-_{k}^{(3)} & J_{1}Z_{1}+J_{3}Z_{3}-J_{2}Z_{2}(\gamma_{k}^{(2)}+\gamma_{-k}^{(2)}-2)
\end{array}\right)\left(\begin{array}{c}
a_{\mathbf{k}}\\
b_{\mathbf{k}}
\end{array}\right).
\label{eq:FM_hamiltonian}
\end{equation}
The energies associated with this Hamiltonian are readily obtained upon diagonalization through a unitary transformation

\begin{equation}
\hbar\omega_{\mathbf{k}}^{\pm}=S(J_{1}Z_{1}+J_{3}Z_{3})+\frac{J_{2}}{2}Z_{2}S\left(2-\gamma_{\mathbf{k}}^{(2)}-\gamma_{-\mathbf{k}}^{(2)}\right)\pm S\sqrt{\left(J_{1}Z_{1}\gamma_{-\mathbf{k}}^{(1)}+J_{3}Z_{3}\gamma_{-\mathbf{k}}^{(3)}\right)\left(J_{1}Z_{1}\gamma_{\mathbf{k}}^{(1)}+J_{3}Z_{3}\gamma_{\mathbf{k}}^{(3)}\right)}.
\end{equation}
\subsection{Antiferromagnet A-AFM Cr$_{2}$CF$_{2}$ }

For the A-AFM (Cr$_{2}$CF$_{2}$) we need to redefine the Holstein-Primakoff transformation as follows

\begin{align}
S_{+}^{A}(\mathbf{R_{i}^{A}})=S_{i,+}^{A}&=\sqrt{2S}\left(1-\frac{a_{i}^{+}a_{i}}{2S}\right)^{\frac{1}{2}}a_{i}\approx\sqrt{2S}a_{i},\\
S_{-}^{A}(\mathbf{R_{i}^{A}})=S_{i,-}^{A}&=\sqrt{2S}\left(1-\frac{a_{i}^{+}a_{i}}{2S}\right)^{\frac{1}{2}}a_{i}^{+}\approx\sqrt{2S}a_{i}^{+},\\
S_{z}^{A}(\mathbf{R_{i}^{A}})=S_{i,z}^{A}&=S-a_{i}^{+}a_{i},
\end{align}
and for the sublattice B
\begin{align}
S_{+}^{B}(\mathbf{R_{i}^{B}})=S_{i,+}^{B}&=\sqrt{2S}\left(1-\frac{b_{i}^{+}b_{i}}{2S}\right)^{\frac{1}{2}}b_{i}^{+}\approx\sqrt{2S}b_{i}^{+},\\
S_{-}^{B}(\mathbf{R_{i}^{B}})=S_{i,-}^{B}&=\sqrt{2S}\left(1-\frac{b_{i}^{+}b_{i}}{2S}\right)^{\frac{1}{2}}b_{i}\approx\sqrt{2S}b_{i}\\
S_{z}^{B}(\mathbf{R_{i}^{B}})=S_{i,z}^{B}&=-S+b_{i}^{+}b_{i},
\end{align}
The difference in definition is due to the change of the spin orientation in the sublattices A and B. Now, the Hamiltonian for the first, second and third neighbors reads

\begin{equation}
{\cal H}_{1}=-J_{1}\sum_{\left\langle i,j\right\rangle }\textbf{S}\left(\mathbf{R}_{i}\right)\cdot\textbf{S}\left(\mathbf{R}_{j}\right)\approx J_{1}S^{2}N{\cal Z}_{1}-{\cal Z}_{1}J_{1}S\sum_{\mathbf{k}}\left(a_{\mathbf{k}}^{\dagger}\quad b_{\mathbf{k}}\right)\left(\begin{array}{cc}
1 & \gamma_{\mathbf{k}}^{\left(1\right)}\\
\gamma_{\mathbf{-k}}^{\left(1\right)} & 1
\end{array}\right)\left(\begin{array}{c}
a_{\mathbf{k}}\\
b_{\mathbf{k}}^{\dagger}
\end{array}\right),
\end{equation}

for the second neighbors, 

\begin{align}
{\cal H}_{2}=-J_{2}\sum_{\left\langle \left\langle i,j\right\rangle \right\rangle }\textbf{S}\left(\mathbf{R}_{i}\right)\cdot\textbf{S}\left(\mathbf{R}_{j}\right) & \approx-J_{2}S^{2}N{\cal Z}_{2}\\
 & +{\cal Z}_{2}\frac{J_{2}}{2}S\sum_{\mathbf{k}}\left(a_{\mathbf{k}}^{\dagger}\quad b_{\mathbf{k}}\right)\left(\begin{array}{cc}
2-\gamma_{\mathbf{-k}}^{\left(2\right)}-\gamma_{\mathbf{k}}^{\left(2\right)} & 0\\
0 & 2-\gamma_{\mathbf{-k}}^{\left(2\right)}-\gamma_{\mathbf{k}}^{\left(2\right)}
\end{array}\right)\left(\begin{array}{c}
a_{\mathbf{k}}\\
b_{\mathbf{k}}^{\dagger}
\end{array}\right)\nonumber,
\end{align}
and finally, for the third neighbors
\begin{equation}
{\cal H}_{3}=-J_{3}\sum_{\left\langle \left\langle \left\langle i,j\right\rangle \right\rangle \right\rangle  }\textbf{S}\left(\mathbf{R}_{i}\right)\cdot\textbf{S}\left(\mathbf{R}_{j}\right)\approx J_{3}S^{2}N{\cal Z}_{3}-{\cal Z}_{3}J_{3}S\sum_{\mathbf{k}}\left(a_{\mathbf{k}}^{\dagger}\quad b_{\mathbf{k}}^{\dagger}\right)\left(\begin{array}{cc}
1 & \gamma_{\mathbf{k}}^{\left(3\right)}\\
\gamma_{\mathbf{-k}}^{\left(3\right)} & 1
\end{array}\right)\left(\begin{array}{c}
a_{\mathbf{k}}\\
b_{\mathbf{k}}
\end{array}\right).
\end{equation}
With final Hamiltonian given by

\begin{equation}
    H=S\sum_{\mathbf{k}}\left(\begin{array}{cc}
a_{\mathbf{k}}^{+} & b_{\mathbf{k}}\end{array}\right)\left(\begin{array}{cc}
-Z_{1}J_{1}+Z_{3}J_{3}+\frac{J_{2}}{2}Z_{2}\left(2-\gamma_{\mathbf{k}}^{(2)}-\gamma_{-\mathbf{k}}^{(2)}\right) & -J_{1}Z_{1}\gamma_{\mathbf{k}}^{(1)}+J_{3}Z_{3}\gamma_{\mathbf{k}}^{(3)}\\
-J_{1}Z_{1}\gamma_{-\mathbf{k}}^{(1)}+J_{3}Z_{3}\gamma_{\mathbf{-k}}^{(3)} & -Z_{1}J_{1}+Z_{3}J_{3}+\frac{J_{2}}{2}Z_{2}\left(2-\gamma_{\mathbf{k}}^{(2)}-\gamma_{-\mathbf{k}}^{(2)}\right)
\end{array}\right)\left(\begin{array}{c}
a_{\mathbf{k}}\\
b_{\mathbf{k}}^{+}
\end{array}\right),
\end{equation}
where again $\gamma_{\pm\mathbf{k}}^{(\nu)}=\frac{1}{Z_{\nu}}\sum_{j}e^{\pm i\mathbf{k}\cdot\mathbf{\mathbf{\delta}_{j}^{(\nu)}}}. \delta_{j}^{(\nu)}$ is the vector connecting the first ($\nu$=1), second ($\nu$=2) and third ($\nu$=3) neighbors. Z$_{\nu}$=3,6,3 for $\nu$=1,2,3 respectively, is the number of neighbors. In spit of the resemblance with the Hamiltonian \ref{eq:FM_hamiltonian}, the A-AFM Hamiltonian must be diagonalized using a paraunitary transformation, which in this simple case is known as Bogoliobov transformation. Defining
\begin{equation}
    H=\sum_{\mathbf{k}}\left(\begin{array}{cc}
a_{\mathbf{k}}^{+} & b_{\mathbf{k}}\end{array}\right)\left(\begin{array}{cc}
A & B\\
B^{*} & A
\end{array}\right)\left(\begin{array}{c}
a_{\mathbf{k}}\\
b_{\mathbf{k}}^{+}
\end{array}\right),
\end{equation}
where
\begin{align}
A &=-S\left(Z_{1}J_{1}+Z_{3}J_{3}\right)+\frac{J_{2}}{2}SZ_{2}\left(2-\gamma_{\mathbf{k}}^{(2)}-\gamma_{-\mathbf{k}}^{(2)}\right), \\
B&= -S\left(J_{1}Z_{1}\gamma_{\mathbf{k}}^{(1)}+J_{3}Z_{3}\gamma_{\mathbf{k}}^{(3)}\right).
\end{align}
Now, performing a transformation for the new basis 
\begin{equation}
    \left(\begin{array}{c}
\alpha_{\mathbf{k}}\\
\beta_{\mathbf{k}}^{+}
\end{array}\right)=\left(\begin{array}{cc}
u_{\mathbf{k}} & v_{\mathbf{k}}\\
v_{\mathbf{k}}^{*} & u_{\mathbf{k}}
\end{array}\right)\left(\begin{array}{c}
a_{\mathbf{k}}\\
b_{\mathbf{k}}^{+}
\end{array}\right),
\end{equation}
and the inverse transformation reads
\begin{equation}
    \left(\begin{array}{c}
a_{\mathbf{k}}\\
b_{\mathbf{k}}^{+}
\end{array}\right)=\left(\begin{array}{cc}
u_{\mathbf{k}} & -v_{\mathbf{k}}\\
-v_{\mathbf{k}}^{*} & u_{\mathbf{k}}
\end{array}\right)\left(\begin{array}{c}
\alpha_{\mathbf{k}}\\
\beta_{\mathbf{k}}^{+}
\end{array}\right).
\end{equation}
Where we have chosen $u_{\mathbf{k}}$ to be real. The new operators must satisfy the same commutation relation as the old basis, therefore,
\begin{align}
    \left[\alpha_{\mathbf{k}},\alpha_{\mathbf{k}}^{+}\right]&=\left[\beta_{\mathbf{k}},\beta_{\mathbf{k}}^{+}\right]=1,\\
    \left[\alpha_{\mathbf{k}},\alpha_{\mathbf{k}}\right]&=\left[\beta_{\mathbf{k}},\beta_{\mathbf{k}}\right]=\left[\alpha_{\mathbf{k}},\beta_{\mathbf{k}}\right]=0.
\end{align}
The relations between $u_{\mathbf{k}}$ and $v_{\mathbf{k}}$ comes from
\begin{align}
    [a_{\mathbf{k}},a_{\mathbf{k}}^{+}]	&=\left[\left(u_{\mathbf{k}}\alpha_{\mathbf{k}}-v_{\mathbf{k}}\beta_{\mathbf{k}}^{+}\right),\left(u_{\mathbf{k}}\alpha_{\mathbf{k}}^{+}-v_{\mathbf{k}}^{*}\beta_{\mathbf{k}}\right)\right],\\
	&=u_{\mathbf{k}}^{2}\left[\alpha_{\mathbf{k}},\alpha_{\mathbf{k}}^{+}\right]-|v_{\mathbf{k}}|^{2}\left[\beta_{\mathbf{k}},\beta_{\mathbf{k}}^{+}\right],\\
	&=u_{\mathbf{k}}^{2}-|v_{\mathbf{k}}|^{2}=1.
\end{align}
Setting 
\begin{align}
    u_{\mathbf{k}}&=\cosh(\theta),\\
    v_{\mathbf{k}}&=\exp(i\phi)\sinh(\theta).
\end{align}
Applying this into the Hamiltonian and requiring that the off-diagonal terms goes to zero, one finds
\begin{equation}
    B^{*}u_{\mathbf{k}}^{2}+Bv_{\mathbf{k}}^{2}-2Au_{\mathbf{k}}v_{\mathbf{k}}=0.
\end{equation}
Solving this quadratic equation for $u_{\mathbf{k}}$ and choosing the phase of $v_{\mathbf{k}}$ in such a way that it cancels that of B, one obtains
\begin{equation}
    \tanh(2\theta)=\frac{|B|}{A}.
\end{equation}
Solving for $\cosh(2\theta)$ and $\sinh(2\theta)$, 
\begin{align}
    \cosh(2\theta)&=\frac{A}{\sqrt{A^{2}-|B|^{2}}},\\\sinh(2\theta)&=\frac{|B|}{\sqrt{A^{2}-|B|^{2}}},
\end{align}
one can write the energies as 
\begin{equation}
    \hbar\omega_{\mathbf{k}}^{\pm}=\sqrt{A^{2}-|B|^{2}}.
\end{equation}
\begin{table}[ht!]
\centering
\caption{Optimized lattice parameters of the MAX phase Cr$_2$AlC, bare and functionalized Cr$_2$C. Here, E$_c$ and E$_f$ represent the calculated values of cohesive and formation energies. The negative values of the cohesive and formation energies reveal their chemical stability. A small positive formation energy in Cr$_2$C indicates that it may require some extra energy for its synthesis. However, it is an experimentally synthesized [3] material. In addition, it is also mechanically stable. The U (= 3 eV) increases the band gap, preserving the indirect gap. The SOC also slightly inferences the band gap of the MXenes. Moreover, we also performed hybrid functional calculations (HSE06) to confirm the results obtained from GGA and GGA + U.}
\vspace*{-0.3cm}
\begin{center}
\begin{tabular}{c c c c c c c c}
\toprule
{Systems} & a (\AA) & E$_\mathrm{c}$ & E$_\mathrm{f}$ & \multicolumn{4}{p{5cm}}{\centering Band gap (eV)}\\ 
\cline{5-8} &{GGA} & {GGA} & {GGA} & GGA & GGA + U & GGA + U + SOC & {HSE06}\\ \hline
					
Cr$_2$AlC &  2.841, 12.681 [4, 5] & -40.461 & -3.371 & metallic & metallic &  metallic & metallic\\
Cr$_2$C &   2.817 [6] & -13.803  & 1.397 & metallic & metallic & metallic & metallic\\
Cr$_2$CF$_2$ &  3.036  & -25.215 & -7.254 [7] & 1.140  &  1.867  & 1.864 & 3.282 \\
Cr$_2$CO$_2$ &   2.893 [6] & -27.352 & -9.139 [7] & metallic & metallic & metallic &  0.160 \\
Cr$_2$CH$_2$ & 2.841 [6] & -19.990  & -0.237  & metallic & 0.422 & 0.380 & 1.725 \\
Cr$_2$C(OH)$_2$ &  3.032 & -32.716 & -9.951 [7] & 0.150   &  0.694 & 0.697 &  1.641\\
Cr$_2$CCl$_2$ & 3.121 & -21.750 & -3.562 & 0.842  &  1.144 & 1.141 & 2.499\\
\hline
\hline
\end{tabular}
\end{center}
\label{T}
\end{table}

\begin{table}[ht!]
\centering
\caption{Calculated elastic stiffness constants (C$_{ij}$), Young's modulus (Y), Shear modulus (G), and Poisson's ratio ($\nu$) of the bare and functionalized Cr$_2$C. The mechanical properties are analyzed by using the VASPKIT [8] post processing tool.}
\vspace*{0.3cm}
\begin{tabular}{l c c c c c c}
\toprule
Systems & C$_{11}$ (Gpa)& C$_{12}$ (Gpa)& Y (Gpa) & $\nu$ & G (Gpa)\\
\hline
Cr$_2$C & 173.79 & 21.46   & 171.14  &   0.12 & 76.17 \\
Cr$_2$CF$_2$ & 151.44  & 101.85 & 82.94 & 0.67 & 24.79 \\
Cr$_2$CO$_2$ & 110.51 & -67.13 & 69.73 & -0.61 & 88.82 \\
Cr$_2$CH$_2$ & 125.19 & -102.37  & 41.48 & -0.82 & 113.78 \\
Cr$_2$C(OH)$_2$ & 208.55  & 61.85 & 190.21 & 0.30 & 73.35\\
Cr$_2$CCl$_2$ &  200.06 & 57.96 & 183.26 & 0.29 & 71.05 \\
\hline
\hline
\end{tabular}
\label{table2}	
\end{table}

\begin{figure*}[!ht]
\centering	
\includegraphics[width=0.99\textwidth]{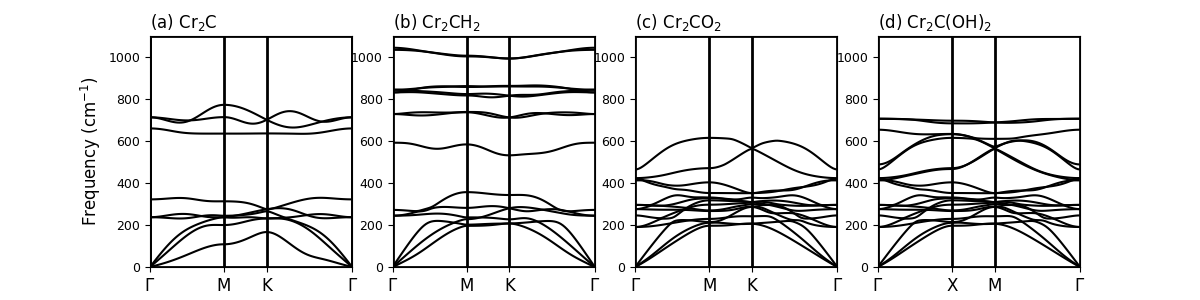}\hfill	
\caption[dop]{Calculated phonon dispersion of the bare and -H, -O, and -OH functionalized Cr$_2$C.}
\label{figure1}
\end{figure*}

\begin{figure*}[!ht]
\centering	
\includegraphics[width=0.95\textwidth]{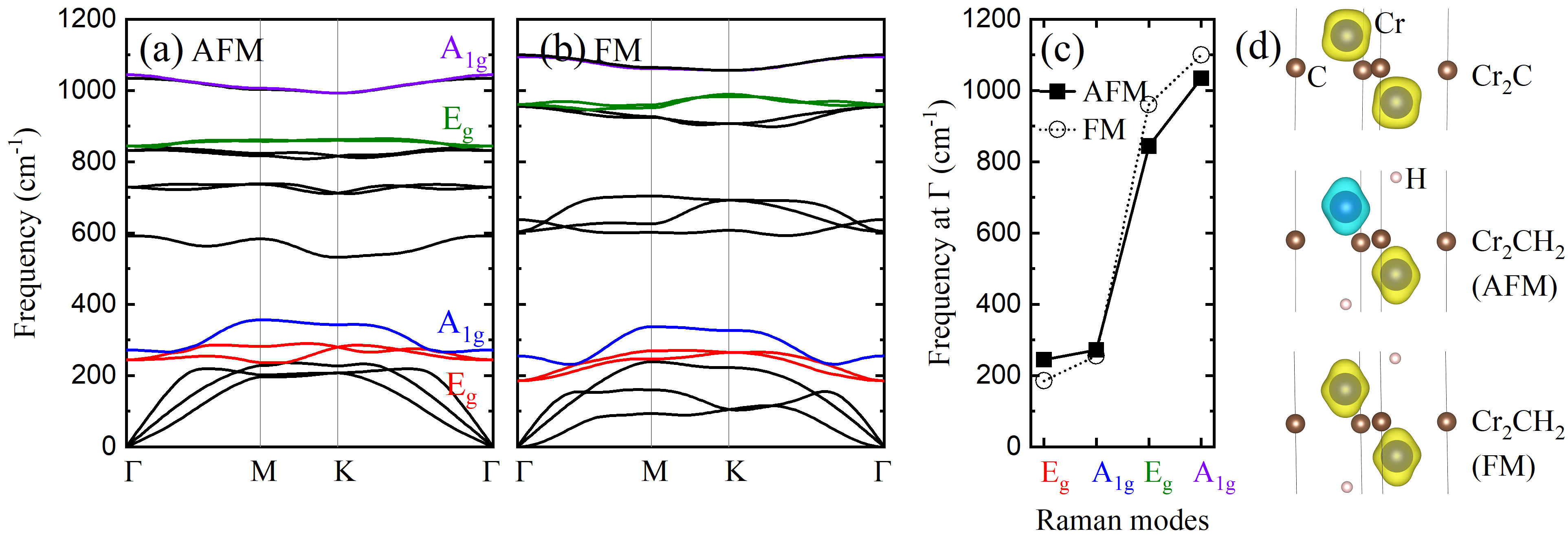}\hfill		
\caption[dop]{Calculated phonon dispersion of Cr$_2$CH$_2$ in (a) AFM and (b) FM spin alignment. Phonon frequencies of the Raman active modes at the $\Gamma$ point (c). Spin density of bare Cr$_2$C and Cr$_2$CH$_2$ (d).}
\label{spin-ph}
\end{figure*}

\begin{table*}[ht!]
\centering
\caption{Calculated defect formation energy values (eV) of the Fe and V doped bare and functionalized Cr$_2$C in their corresponding ground states. Here, 2 $\times$ 2 $\times$ 1 supercell is used to model the AFM structure. The Fe doped systems are less stable as compared to the V doped under the GGA. However, the GGA + U makes it chemically stable. Thus, the on site correlation effect plays a significant role to stabilize the system, especially for the Fe doped ones.}
\vspace*{0.3cm}
\begin{tabular}{l c  c | c c  c c c }
\toprule
Systems &\multicolumn{2}{p{5cm}} {\centering GGA} &\multicolumn{4}{p{5cm}} {\centering GGA + U} \\ \cline{2-5} &{Fe-doped (eV)} & {V-doped (eV)}  &{Fe-doped (eV)} & {V-doped (eV)} \\
\hline
Cr$_2$ C & -1.14 (FM) & -2.01 (C-AFM) & -2.85 (G-AFM) & -0.86 C-AFM\\
Cr$_2$CF$_2$ & 0.24 (A-AFM)  & -1.79 (A-AFM) & -0.54 (A-AFM)  &  -1.43 (A-AFM) \\
Cr$_2$CO$_2$ & -0.06 (FM) & -2.68 (G-AFM) & -0.10 (FM) & -2.39 (FM) \\
Cr$_2$CH$_2$ & -0.60 (G-AFM) & -1.95 (C-AFM)  &  -0.88 (C-AFM) &  -1.26 (C-AFM) \\
Cr$_2$C(OH)$_2$ & 0.14 (A-AFM) & -1.87 (A-AFM) &  -0.62 (A-AFM) &  -1.34 (A-AFM) \\
Cr$_2$CCl$_2$ & 0.10 (A-AFM) & -1.79 (G-AFM) &  -0.87 (A-AFM) &  -1.48 (A-AFM) \\
\hline
\hline
\end{tabular}
\label{table3}	
\end{table*}

\begin{table*}[ht!]
\centering
\caption{Calculated elastic stiffness constants (C$_{ij}$), Young's modulus (Y), Shear Modulus (G), and Poisson's ratio ($\nu$) of the Fe and V doped bare and functionalized Cr$_2$C.} 
\vspace*{0.3cm}
\begin{tabular}{l c c c c c | c c c c c c c }
\toprule
Systems &\multicolumn{5}{p{4cm}} {\centering Fe doping} &\multicolumn{5}{p{5cm}} {\centering V doping} \\
\cline{2-11} & C$_{11}$ &  C$_{12}$  & Y & $\nu$ & G &C$_{11}$ &  C$_{12}$  & Y & $\nu$ & G \\
\hline
Cr$_2$C & 106.93 & -93.69 & 24.84 & -0.88 & 100.31 & 161.12 & 5.24 & 160.95 & 0.03 &  77.94\\
Cr$_2$CF$_2$ & 181.63  & 57.93 &  163.15  &  0.32  & 61.85 & 167.70 & 77.46 & 131.92 & 0.46 & 45.12\\
Cr$_2$CO$_2$ & 789.00 & -495.61 & 477.68 & -0.63 & 642.30 &372.32 &-310.12 & 114.00 &-0.83 & 341.22 \\
Cr$_2$CH$_2$ &  190.38 & 0.60 & 190.38 &  0.00 & 94.89 & 177.47 & 62.26 & 155.63 &  0.35 & 57.61 \\
Cr$_2$C(OH)$_2$ &  143.15  & 96.24  & 78.46  &  0.67 & 23.46 &  176.97 &68.90 &150.15 & 0.39  & 54.04  \\
Cr$_2$CCl$_2$ & 183.38 & 53.17 & 167.96 &0.29 &  65.10 & 175.80 & 53.35  & 159.61 & 0.30 & 61.23 \\
\hline
\hline
\end{tabular}
\label{table4}	
\end{table*}

\begin{table*}[ht!]
\centering
\caption{Calculated electronic properties of Fe and V doped bare and functionalized Cr$_2$C with different approaches, such as GGA, GGA + U, and HSE06. Here, M and HM indicate metallic and half metallic, respectively.}
\vspace*{0.3cm}
\begin{tabular}{c | c  c | c  c |c  c }
\toprule
Systems &\multicolumn{2}{ p{3cm}} {\centering GGA (eV)} &\multicolumn{2}{p{3cm}} {\centering GGA + U (eV} & \multicolumn{2}{p{3cm}} {\centering HSE06 (eV)}\\
\cline{2-7} &{Fe} & {V }  &{Fe} & {V} & Fe & V\\
\hline
Cr$_2$C & M & M & M & M   & .. & ..\\
Cr$_2$CF$_2$ &  0.46  & HM & 1.63 &  1.48 & .. &..\\
Cr$_2$CO$_2$ & HM & M & HM & M & .. & .. \\
Cr$_2$CH$_2$ & M & M &  0.29 &  0.28 & .. & .. \\
Cr$_2$C(OH)$_2$ &  0.04 & HM &  0.26 & 0.80 & .. & 1.31\\
Cr$_2$CCl$_2$ & 0.37 & 0.18 &  0.27 & 0.86 & 1.80 &  1.64 \\
\hline
\hline
\end{tabular}
\label{table4}	
\end{table*}

\begin{figure*}[!ht]
\centering	
\includegraphics[width=0.99\textwidth]{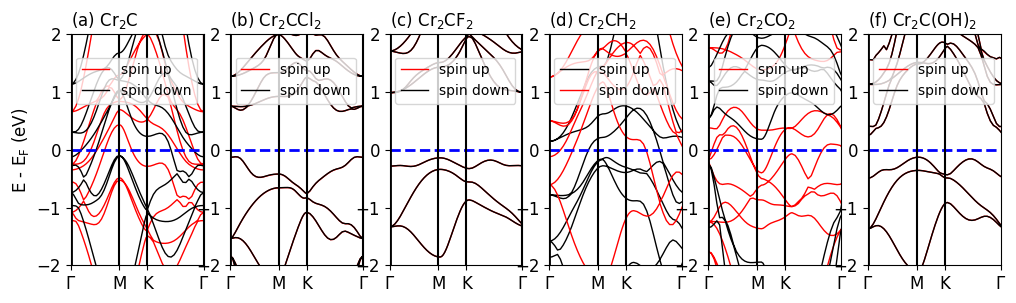}\hfill	
\caption[dop]{Calculated band structure of the (a) bare and (b)-(f) functionalized Cr$_2$C with the GGA.}
\label{figure1}
\end{figure*}

\begin{figure*}[!ht]
\centering	
\includegraphics[width=0.99\textwidth]{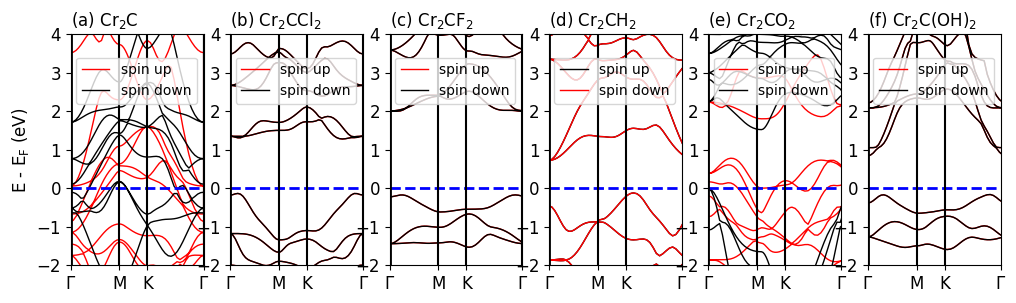}\hfill	
\caption[dop]{Calculated band structure of the (a) bare and (b)-(f) functionalized Cr$_2$C with the GGA + U.}
\label{figure1}
\end{figure*}

\begin{figure*}[!ht]
\centering	
\includegraphics[width=0.99\textwidth]{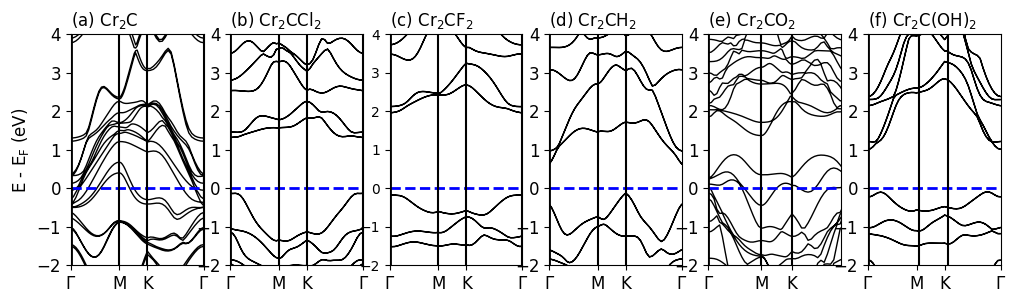}\hfill	
\caption[dop]{Calculated band structure of the (a) bare and (b)-(f) functionalized Cr$_2$C with the GGA + U including the SOC effect.}
\label{figure1}
\end{figure*}

\begin{figure*}[!ht]
\centering	
\includegraphics[width=0.99\textwidth]{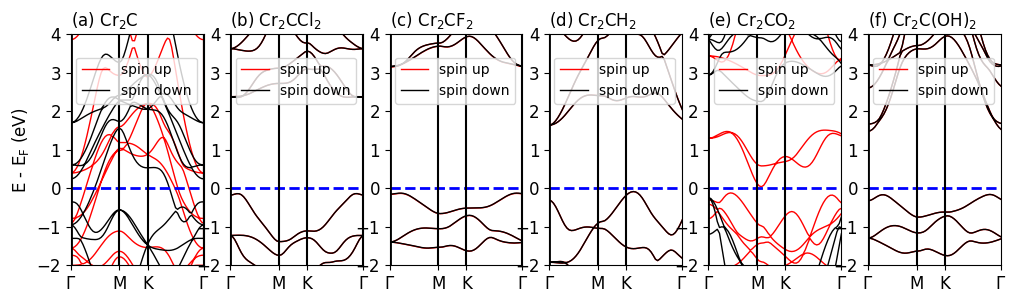}\hfill	
\caption[dop]{Calculated band structure of the (a) bare and (b)-(f) functionalized Cr$_2$C with the HSE06.}
\label{figure2}
\end{figure*}

\begin{figure*}[!ht]
\centering	
\includegraphics[width=0.99\textwidth]{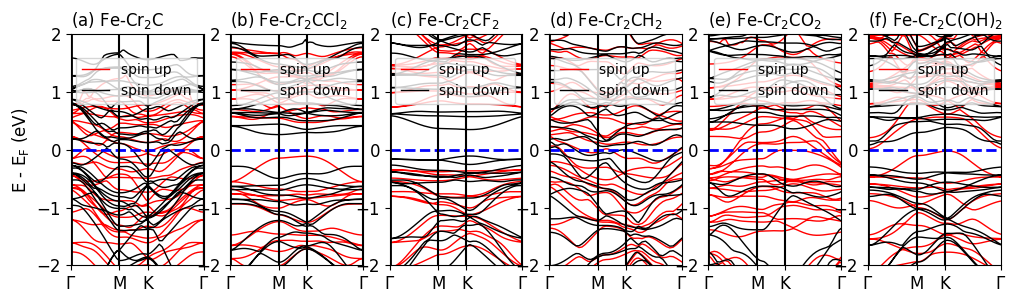}\hfill	
\caption[dop]{Calculated band structure of the Fe doped (a) bare and (b)-(f) functionalized Cr$_2$C with the GGA.}
\label{figure3}
\end{figure*}

\begin{figure*}[!ht]
\centering	
\includegraphics[width=0.99\textwidth]{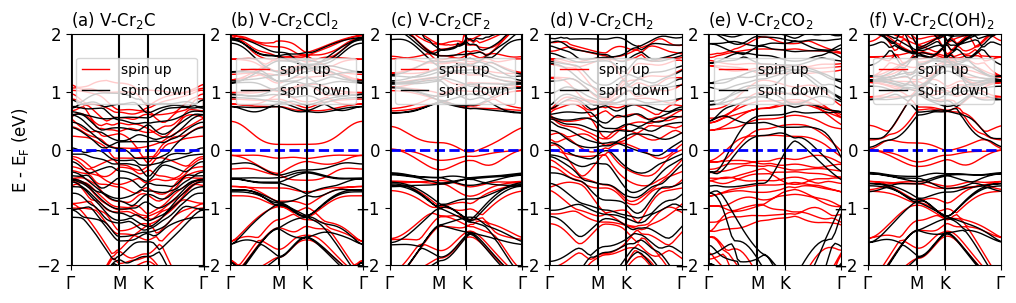}\hfill	
\caption[dop]{Calculated band structure of the V doped (a) bare and (b)-(f) functionalized Cr$_2$C with the GGA.}
\label{figure4}
\end{figure*}
\begin{figure*}[!ht]
\centering	
\includegraphics[width=0.8\textwidth]{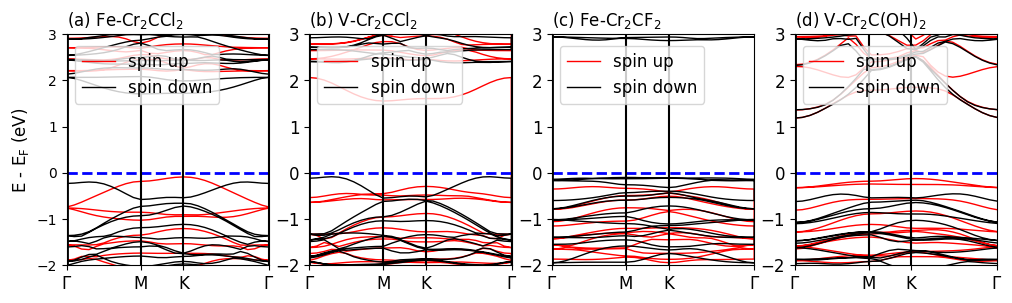}\hfill	
\caption[dop]{Calculated band structure of (a) the Fe doped Cr$_2$CCl$_2$, (b) the V doped Cr$_2$CCl$_2$, (c) the Fe doped Cr$_2$CF$_2$, and (d) the V doped Cr$_2$C(OH)$_2$ with the HSE06.}
\label{figure7}
\end{figure*}
\begin{figure*}[!ht]
\centering	
\includegraphics[width=0.99\textwidth]{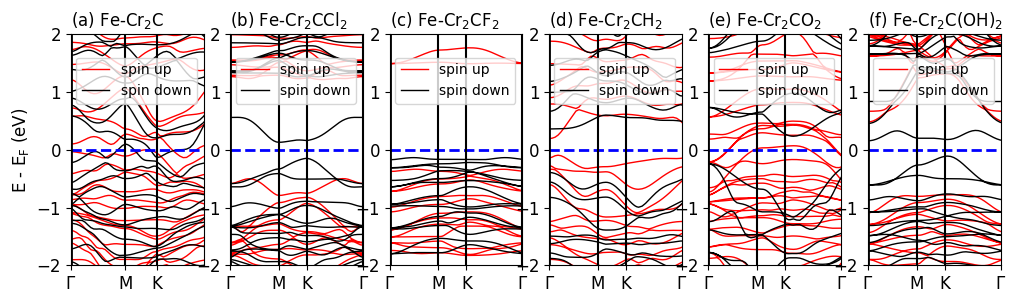}\hfill	
\caption[dop]{Calculated band structure of the Fe doped (a) bare and (b)-(f) functionalized Cr$_2$C with the GGA + U.}
\label{figure3}
\end{figure*}

\begin{figure*}[!ht]
\centering	
\includegraphics[width=0.99\textwidth]{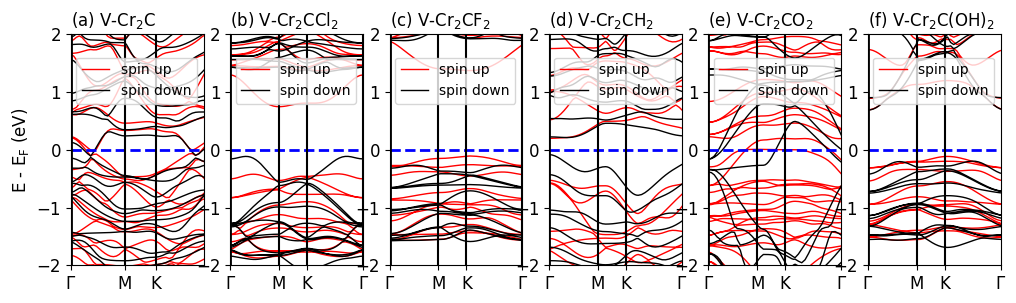}\hfill	
\caption[dop]{Calculated band structure of the V doped (a) bare and (b)-(f) functionalized Cr$_2$C with the GGA + U.}
\label{figure6}
\end{figure*}

\begin{table}[ht!]
\centering
\caption{Calculated magnetic moments on individual Cr atom of the bare and functionalized Cr$_2$C with the GGA, GGA + U, and HSE06. The calculations are performed using a unit cell structure.}
\vspace*{0.3cm}
\begin{tabular}{l c c c c c }
\toprule
Systems & GGA ($\mu_B$) & GGA + U ($\mu_B$) & HSE06 ($\mu_B$)\\
\hline
Cr$_2$C & 0.50  & 1.00 & 0.83\\
Cr$_2$CF$_2$ & 2.54  &  3.08 & 2.97 \\
Cr$_2$CO$_2$ & 2.09 & 2.47  & 2.43 \\
Cr$_2$CH$_2$ & 1.48 & 3.02 & 2.90\\
Cr$_2$C(OH)$_2$ & 2.40 & 3.00 & 2.89\\
Cr$_2$CCl$_2$ & 2.21 & 2.90 & 2.82 \\
\hline
\hline
\end{tabular}
\label{table5}	
\end{table}

\begin{table}[ht!]
\centering
\caption{Calculated exchange coupling parameters J$_1$, J$_2$, and J$_3$ with the GGA. These parameters are calculated by using 2 $\times$ 2 $\times$ 1 supercell structure.} 
\begin{tabular}{l  c c c c}
\toprule
Systems & \multicolumn {4}{p{6cm}} {\centering {GGA}} \\ 
\cline{2-5} & J$_{1}$ & J$_{2}$ & J$_{3}$  & stable configurations \\
\hline
Cr$_2$C & 1.93 & -0.60  & -0.45 & C-AFM \\
Cr$_2$CF$_2$ &  -48.48 &  21.54 & -5.25 & A-AFM\\
Cr$_2$CO$_2$ & 4.04 & 2.38 &   0.73 &  FM \\
Cr$_2$CH$_2$ & -1.67 & -10.65 &  -1.21  &  C-AFM  \\

Cr$_2$C(OH)$_2$ & -49.54 & 14.49 & -10.64 & A-AFM \\
Cr$_2$CCl$_2$ & -32.83 & 18.34 &  -8.63  &  A-AFM  \\

\hline
\hline
\end{tabular}
\label{table6}	
\end{table}

\begin{table*}[ht!]
\centering
\caption{Calculated magnetic moments on each magnetic ions (Cr, Fe, and V) with the GGA and GGA + U (or HSE06) of the Fe and V doped bare and functionalized Cr$_2$C.}
\vspace*{0.3cm}
\begin{tabular}{l c c c c c c c  c | c c c c c c c c}
\toprule
Systems &\multicolumn{7}{p{4cm}} {\centering GGA} &\multicolumn{8}{p{5cm}} {\centering GGA + U/HES06} \\
\cline{2-17} & Cr$_{1}$ &  Cr$_{2}$  & Cr$_{3}$ & Cr$_{4}$ & Cr$_{5}$ & Cr$_{6}$ & Cr$_{7}$ & Fe/V&Cr$_{1}$ &  Cr$_{2}$  & Cr$_{3}$ & Cr$_{4}$& Cr$_{5}$ &Cr$_{6}$ &Cr$_{7}$ & Fe/V \\
\hline
Fe:Cr$_2$C & 0.09& 0.09 &  1.56 & 0.09 & 0.73 & 0.73 & 0.73&   1.83 & -2.93 & 3.07 & -3.35 &  3.07 &  3.04& 3.04 & -3.21 & -2.73\\
V:Cr$_2$C & 0.03& -1.29 &  1.24 & -1.29 & 0.45 & 0.45 & -1.74 &   -0.04 &  3.16 & -3.23 & 3.11 & -3.23 & 3.13& 3.13 & -3.35 & -1.89\\
Fe:Cr$_2$CCl$_2$ & -2.24& -2.24 & -1.53 & -2.24 & 2.27 & 2.27 & 2.27 & 3.04 & -2.84 & -2.84 & -2.41 & -2.84 &  2.86&  2.86 &  2.86 & 3.68\\
V:Cr$_2$CCl$_2$ & -2.20& -2.20 & -2.14 & -2.20 & 2.21 & 2.21 & 2.21 & 1.07 & -2.83 & -2.83 & -2.78 & -2.83 &  2.83&  2.83 &  2.83& 1.71\\
Fe:Cr$_2$CF$_2$ & -2.48& -2.48 & -2.69 & -2.48 & 2.52 & 2.52 & 2.52 &  0.58 & -2.98 & -2.98 & -3.08 & -2.98&  2.96&  2.96 &  2.96& 1.09\\
V:Cr$_2$CF$_2$ & -2.45& -2.45 & -2.38 & -2.45 & 2.46 & 2.46 & 2.46 &  1.34 & -2.94 & -2.94 & -2.88 & -2.94&  2.93&  2.93 &  2.93& 1.84\\
Fe:Cr$_2$CH$_2$ & 1.80&-1.71 & 1.64& -1.71 & 1.67 & 1.67 & -1.79 & -0.23 & 2.88 & -2.70 & 3.07 & -2.70&  2.89& 2.89 & -2.68& -1.48\\
V:Cr$_2$CH$_2$ & -1.39&  1.57 & -1.31& 1.57& 1.33 & 1.33 & -1.37 & -0.05 & 2.84 & -2.79 & 2.72 & -2.79&  2.77& 2.77 & -2.86& -1.67\\
Fe:Cr$_2$CO$_2$ & 2.09&  2.09 & 0.68& 2.09&  2.02 &  2.02 &  2.02 & 1.47 & 2.47 & 2.47 & 2.32 & 2.47&  2.44& 2.44 & 2.44& -2.46\\
V:Cr$_2$CO$_2$ & 1.96&  1.96 & 0.18& 1.96&  1.98 &  1.98 &  1.98 & 0.15 & 2.48 & -2.23 & 2.27 & -2.23&  2.43& 2.43 & -2.42& 0.08\\
Fe:Cr$_2$C(OH)$_2$ & -2.31&  -2.31 & -1.78& -2.31& 2.40&  2.40 &  2.40 & 3.14 & -2.98 & -2.98 & -2.91 & -2.98&  2.96& 2.96 & 2.96& 1.88\\
V:Cr$_2$C(OH)$_2$ & -2.33&  -2.33 & -2.25& -2.33& 2.34&  2.34 &  2.34 & 1.15 & -2.88 & -2.88 & -2.82 & -2.88&  2.88& 2.88 & 2.88& 1.77\\
\hline
\hline
\end{tabular}
\label{table3}	
\end{table*}

\begin{table*}[!ht]
\centering
\caption{Calculated total magnetic moment per unit cell  under compressive and tensile strains. The positive and negative signs indicate the tensile and compressive strains. The strain calculations were performed with GGA level for Cr$_2$C, Cr$_2$C(OH)$_2$, Cr$_2$CF$_2$, and Cr$_2$CCl$_2$. In addition,  GGA + U and HSE06 calculations were performed for Cr$_2$CH$_2$ and  Cr$_2$CO$_2$.} 
\begin{tabular}{l  c c c c c c}
\toprule
Systems & \multicolumn {6}{p{6cm}} {\centering {GGA/HSE06/GGA + U}} \\ 
		\cline{2-7} & 0 & $\mp$ 2 & $\mp$ 4 & $\mp$ 6  & $\mp$ 8 & $\mp$ 10 \\
		\hline
		Cr$_2$C & 1.02 & 1.19/0.97  & 0.00/0.87 & 0.00/0.73 & 00/0.80 & 00/1.41 \\
		Cr$_2$C(OH)$_2$ & 0.00 & 0.00/0.00  & 0.00/0.00 & 0.02/0.00 & 0.02/0.00 & 0.00/0.00 \\
		Cr$_2$CF$_2$ & 0.00 & 0.00/0.00  & 0.00/0.00 & 0.00/0.00 & 0.93/0.00 & 0.00/0.00 \\
		Cr$_2$CCl$_2$ & 0.00 & 0.00/0.00  & 0.00/0.00 & 0.00/0.00 & 0.79/0.00 & 2.25/0.00 \\
		Cr$_2$O$_2$ & 4.00 & 4.00/4.00  & 2.94/4.00 & 4.00/4.00 & 4.00/4.00 & 4.00/4.00 \\
		Cr$_2$H$_2$ & 0.00 & 0.00/0.00  & 1.46/0.00 & 0.00/0.00 & 1.00/0.00 & 0.10/0.00 \\
		\hline
		\hline
	\end{tabular}
	\label{table9}	
\end{table*}	
\begin{figure*}[!ht]
\centering	
\includegraphics[width=0.99\textwidth]{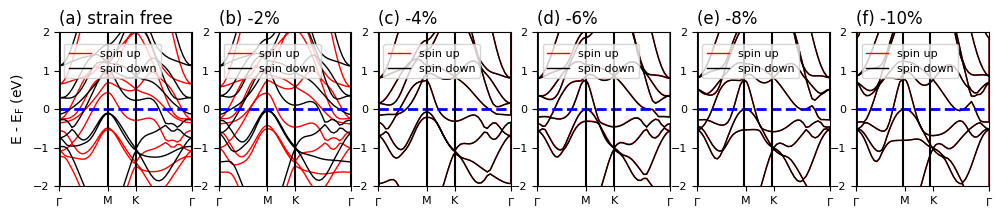}\hfill
\includegraphics[width=0.99\textwidth]{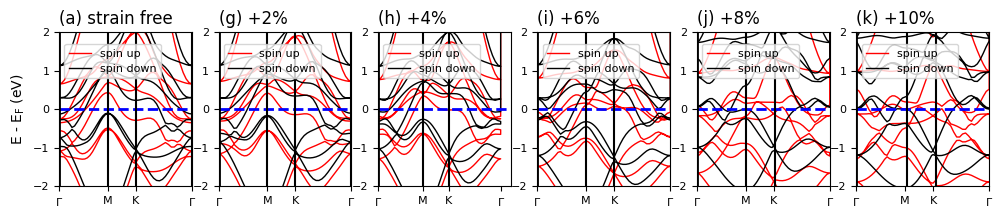}\hfill	
\caption[dop]{Calculated band structure of the bare Cr$_2$C under compressive (negative) and tensile (positive) strains with the GGA.}
\label{figure9}
\end{figure*}

\begin{figure*}[!ht]
\centering	
\includegraphics[width=0.99\textwidth]{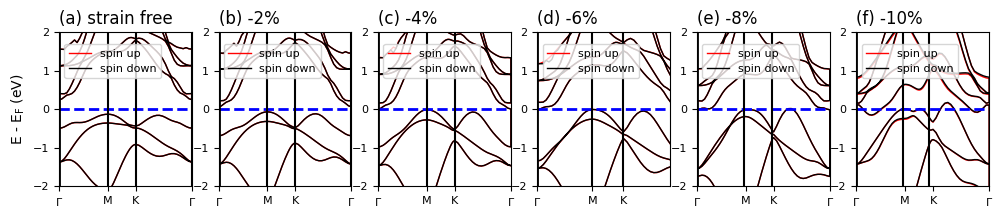}\hfill	
\includegraphics[width=0.99\textwidth]{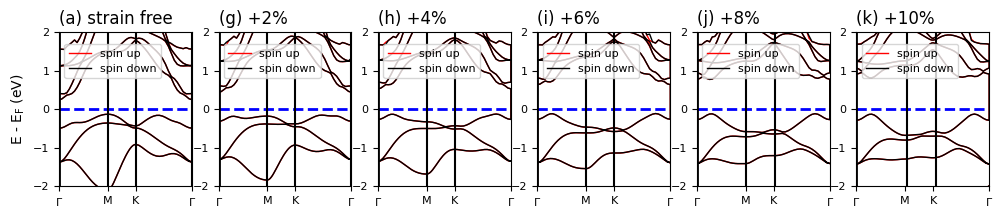}\hfill	
\caption[dop]{Calculated band structure of the Cr$_2$C(OH)$_2$ under compressive (negative) and tensile (positive) strains with the GGA.}
\label{figure14}
\end{figure*}

\begin{figure*}[!ht]
\centering	
\includegraphics[width=0.99\textwidth]{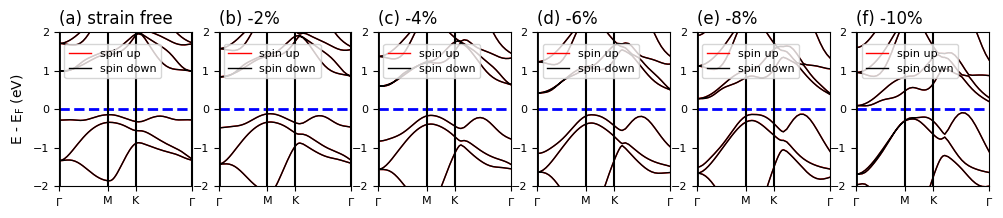}\hfill	
\includegraphics[width=0.99\textwidth]{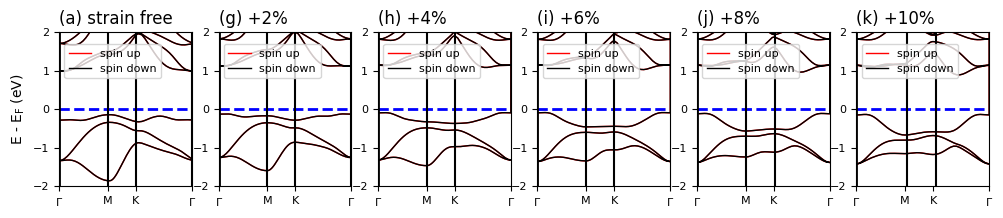}\hfill	
\caption[dop]{Calcualted band structure of Cr$_2$CF$_2$ under compressive (negative) and tensile (positive) strains with the GGA.}
\label{figure12}
\end{figure*}

\begin{figure*}[!ht]
\centering	
\includegraphics[width=0.99\textwidth]{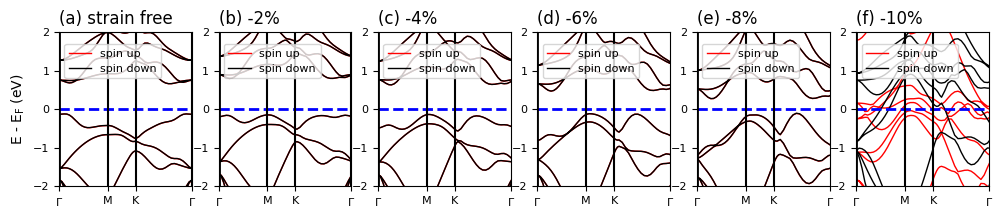}\hfill	
\includegraphics[width=0.99\textwidth]{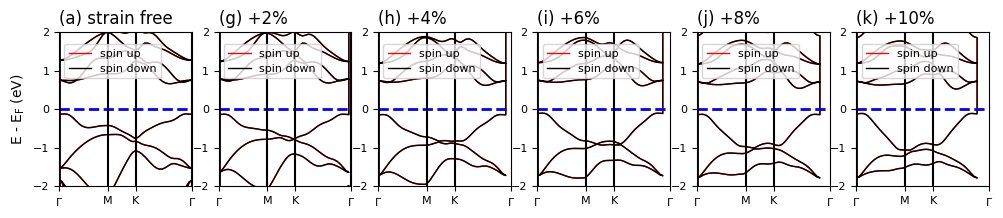}\hfill	
\caption[dop]{Calculated band structure of Cr$_2$CCl$_2$ under compressive (negative) and tensile (positive) strains with the GGA.}
\label{figure11}
\end{figure*}

\begin{figure*}[!ht]
\centering	
\includegraphics[width=0.99\textwidth]{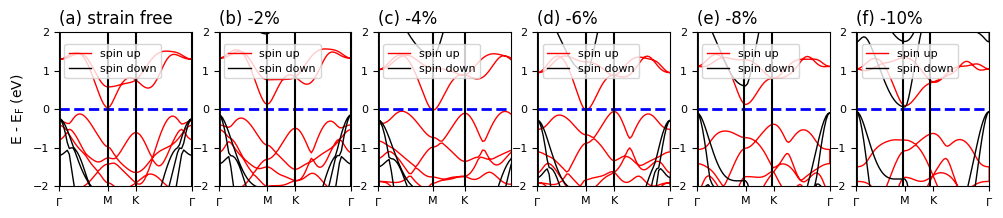}\hfill	
\includegraphics[width=0.99\textwidth]{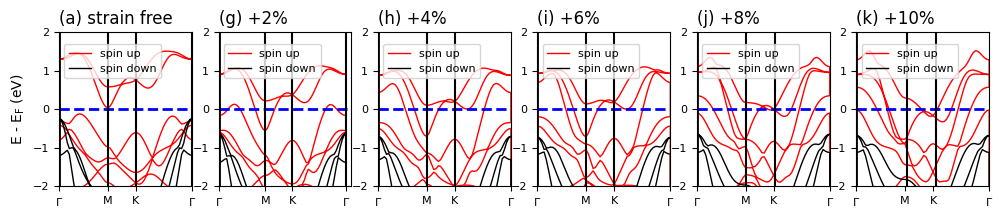}\hfill	
\caption[dop]{Calculated band structure of Cr$_2$CO$_2$ under compressive (negative) and tensile (positive) strains with the HSE06.}
\label{figure16}
\end{figure*}

\begin{figure*}[!ht]
\centering	
\includegraphics[width=0.99\textwidth]{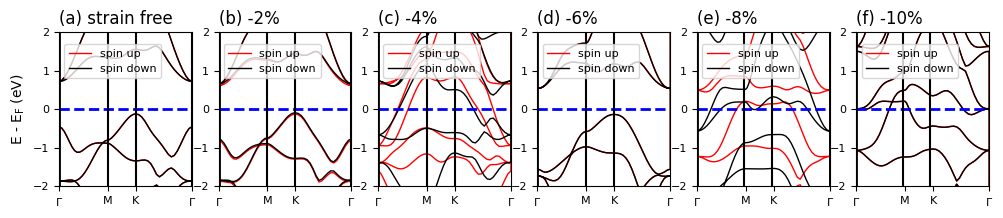}\hfill
\includegraphics[width=0.99\textwidth]{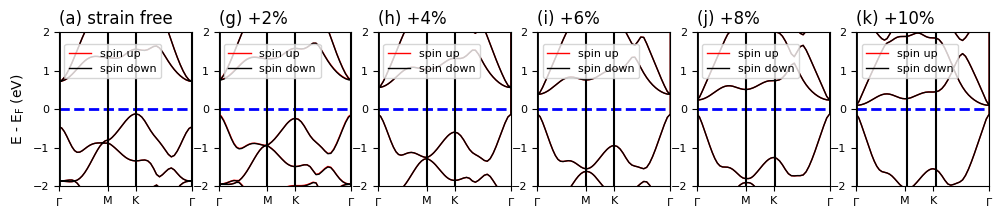}\hfill	
\caption[dop]{Calculated band structure of Cr$_2$CH$_2$ under compressive (negative) and tensile (positive) strains with GGA + U.}
\label{figure18}
\end{figure*}

\begin{figure*}[!ht]
\centering	
\includegraphics[width=0.99\textwidth]{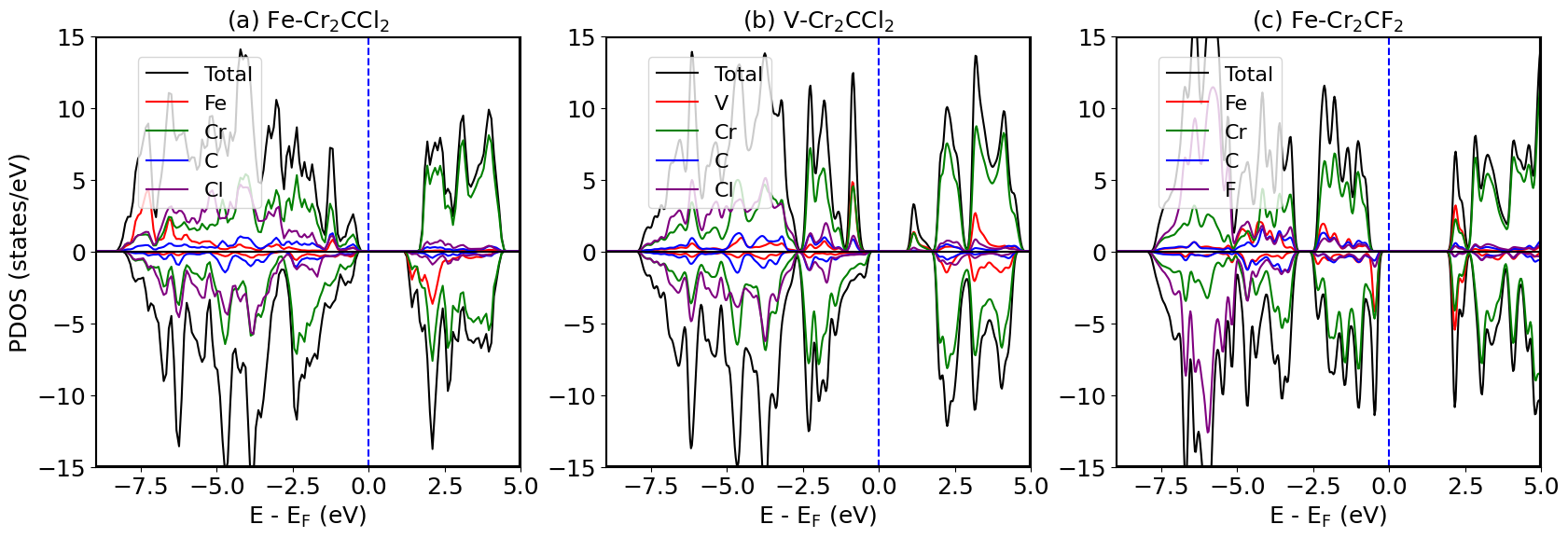}\hfill	
\caption[dop]{Calcualtaed projected density of states (PDOS) of the (a) Fe doped Cr$_2$CCl$_2$, (b) V doped Cr$_2$CCl$_2$, and (c) Fe doped Cr$_2$CF$_2$ with the HSE06.}
\label{figure20}
\end{figure*}
\FloatBarrier
[1] T. Holstein and H. Primakoff, Phys. Rev. 58, 1098 (1940).

[2] A. Prabhakar and D. D. Stancil, “Spin waves: Theory and applications,” (2009).

[3] S. Ma, J. Ye, A. Liu, Y. Liu, J. Wang, and J. Pang, J. Am. Ceram. Soc 99, 1943 (2016).

[4] J. M. Schneider, Z. Sun, R. Mertens, F. Uestel, and R. Ahuja, Solid State Commun. 130, 445 (2004).

[5] S. Cui, D. Wei, H. Hu, W. Feng, and Z. Gong, J. Solid State Chem. 191, 147 (2012).

[6] Y. Wang, J. Xue, G. Nie, and X. Guo, Chem. Phys. Lett. 750, 137444 (2020).

[7] M. Khazaei, M. Arai, T. Sasaki, C.-Y. Chung, N. S. Venkataramanan, M. Estili, Y. Sakka, and Y. Kawazoe, Adv. Funct. Mater. 23, 2185 (2013).

[8] V. Wang, N. Xu, J.-C. Liu, G. Tang, and W.-T. Geng, Comput. Phys. Commun. 267, 108033 (2021).